\documentclass[12pt]{article}
\usepackage{color,amsmath,mathtools,setspace,hyperref,
array,tabu,stackengine,imakeidx,amssymb,graphicx,amsfonts,
xcolor,soul,slashed,subcaption,pgfplots,pdflscape,resizegather}
\usepackage{colortbl}
\usepackage{xcolor}
\usepackage{amsmath,lipsum}
\usepackage{amsfonts}
\usepackage{geometry}
\usepackage{amssymb,tikz,pgfplots}
\usetikzlibrary{shapes,arrows,snakes}
\usepackage[utf8]{inputenc}
\usepackage[english]{babel}
\DeclarePairedDelimiter\bra{\langle}{\rvert}
\DeclarePairedDelimiter\ket{\lvert}{\rangle}

\DeclarePairedDelimiterX\braket[2]{\langle}{\rangle}{#1 \delimsize\vert #2}
\allowdisplaybreaks
\newcommand\beq{\begin{equation}}
\newcommand\eeq{\end{equation}}
\newcommand\bea{\begin{eqnarray}}
\newcommand\eea{\end{eqnarray}}

\begin{document}
\phantom{xxx} 
\vspace{-2.0cm}
\bigskip
\begin{center} 
{\Large \bf Spin Networks, Wilson Loops  and 3nj Wigner Identities
} 
\end{center} 
\vskip .5 true cm
\begin{center} 
{\bf Manu Mathur} \footnote{manu@boson.bose.res.in, manumathur14@gmail.com}, 
{\bf Atul Rathor} \footnote{atulrathor@bose.res.in, atulrathor999@gmail.com}
\vskip 0.6 true cm
S. N. Bose National Centre for Basic Sciences \\ 
JD Block, Sector III, Salt Lake City, Kolkata 700106, India
\end{center} 
\bigskip

\centerline{\bf Abstract}

\noindent
We   exploit the spin network properties of the magnetic eigenstates of SU(2) Hamiltonian lattice gauge theory and use the Wilson loop operators to obtain a wide class of new identities amongst 3nj Wigner coefficients. We also show 
that the topological ground states of the SU(2) toric code Hamiltonian lead to Wigner 3nj identities with non-trivial phases. The method is very general and involves only the  eigenvalue equations of any gauge invariant operator and their solutions. Therefore, it can  be extended to any higher dimensional spin networks as well as larger SU(N) groups. 


\section{\bf Introduction}
\label{intros} 

The idea of spin networks was first introduced by Roger Penrose  as a simple model of discrete quantum space time geometry \cite{pen}. They are now ubiquitous in quantum field theories of elementary particles and gravity  
providing  the most economical as well as complete  description of the their physical  Hilbert spaces.  
In  Hamiltonian lattice gauge theories, the  spin networks solve the Mandelstam constraints present in the loop basis \cite{mand,manuloop}. 
In loop quantum gravity, spin networks represent the quantum states of space geometry \cite{pen,mand,smolin}.   They also play  important roles in  conformal field theories and topological quantum field theories (see references in Lee Smolin in \cite{smolin}). The spin networks are also useful 
to  describe the 
degenerate topological ground states of SU(2) toric code model \cite{manatul} and might be significant 
in topological quantum computing in the future. Infact, in the past there are proposals   to use 
spin networks as storage devices for quantum information in quantum computing \cite{aabbcc}.
In this work, we exploit the group theoretical properties of the spin networks in SU(2) lattice gauge theory  
to obtain a  large class of new identities involving various Wigner 3nj coefficients.
These identities are obtained by applying various Wilson loop operators on the gauge invariant 
magnetic field eigenstates and then computing their action  on  the spin network basis with the help of  generalized Wigner Eckart theorem (see section \ref{wlwi}).
Similar relations have been obtained in the context of quantum gravity in the past \cite{livine}. 
We also use   SU(2) toric code Hamiltonian \cite{manatul} 
on a two dimensional torus as it is exactly solvable and has 4 topological ground states. We show that the  non-contractible Wilson or Polyakov loops encircling the torus in either direction lead to  richer Wigner coefficient identities with topological phases of the ground states 
(see section \ref{tcm}). 
We will work in $d=2,3$ space dimensions and consider 
only simple 1, 2 plaquette contractible or a non-contractible (torus) Wilson loop operators on small lattices. This is to  keep the discussion simple.  More general cases in higher dimensions and with larger  loop operators  can be similarly handled. They  will 
lead to  higher Wigner 3nj coefficients identities.  Further,  
the method can  be generalized to higher SU(N) groups.

The plan of the paper is as follows. In section \ref{intro}, we discuss the kinematical aspects of lattice gauge theory Hamiltonian and summarize the properties and construction of SU(2) spin networks. 
This section is for the sake of completeness and setting up the notations to be used in the following sections. More details about the Hamiltonian formulation of lattice gauge theory  can be found in \cite{kogsus}. The SU(2) spin networks are 
discussed in detail in \cite{mand,manuloop,smolin}. In section \ref{wlwi}, we start with introducing  the basic idea of using the magnetic field eigenvalue equation as generating function for the various identities for 3nj Wigner coefficients.  In section \ref{tetra} and \ref{box}, we consider SU(2) gauge theory on a single tetrahedron and a single cube respectively to illustrate the ideas in the simplest possible settings. Even on these  simple and small 
lattices we get identities involving $6j$ \& $9j$ (tetrahedron case), $12j$ \& $12j$  (cube) Wigner coefficients respectively. 
In section \ref{tcm}, we consider 
the topological ground states of the SU(2) 
toric code Hamiltonian on a small 4 plaquette torus. We show that the identities derived 
 from the contractible Wilson loop operators  are
independent of topological phases but non-contractible 
Wilson loop operators lead to identities decorated with the topological phases of the ground state considered (equation (\ref{mevetp})). This is expected result as only non-local operators on the entire torus can detect topological charges. 
The final results are summarized at the end in the  summary and discussion section. 
 
\section{SU(2) Lattice Gauge Theory and Spin Networks} 
\label{intro} 

In this section we briefly introduce the electric, magnetic field operators, their algebras and  the constraints in the Hamiltonian formulation of lattice gauge theory in $(d+1)$ dimensions. The construction of spin network states and their properties are briefly discussed in section \ref{spinn}.   
We  consider the SU(2) link holonomies $U_{\alpha\beta} (\vec n;\,\hat i\,)$ and their conjugate electric fields $E^a_+(\vec n;\,\hat i\,) ~(E^a_-(\vec n+\hat i;\,\hat i\,))$ on every link $(\vec n;\,\hat i\,)$ ( see Figure -\ref{link}) with $\alpha,\beta =1,2$ and $a=1,2,3$ \cite{kogsus}. 
The electric fields $E^a_+(\vec n;\,\hat i\,)$  $(E^a_-(\vec n+ \hat i;\;\hat i\,))$ rotate (anti-rotate) the link holonomies $U(\vec n;\,\hat i\,)$ from the left (right) and therefore satisfy the following 
canonical commutation relations: 
\bea
[E^{  a}_+(\vec n;\,\hat i\,),~U_{\alpha\beta}(\vec n;\,\hat i\,)]=\left(\frac{\sigma^{a}}{2}U(\vec n;\,\hat i\,)\right)_{\alpha\beta}, ~~[E^{  a}_-(\vec n+\hat i;\,\hat i\,),U_{\alpha\beta}(\vec n;\,\hat i\,)]=-\left(U(\vec n;\,\hat i\,)\frac{\sigma^{a}}{2}\right)_{\alpha\beta}, 
\label{comrel1}
\eea
where  $\sigma^{a}$ are the Pauli matrices. The above commutation relations and Jacobi identity imply 
\bea \label{la}
[E^a_+(n;\,\hat i\,), E^b_+(n;\,\hat i\,)] =\, i\,\epsilon^{abc}~ E^c_+(n;\,\hat i\,),  ~~
[E^a_-(\vec n+ \hat i;\hat i\,),E^b_-(\vec n+ \hat i;\,\hat i\,)]= \,i\,\epsilon^{abc} ~E^c_-(\vec n+ \hat i;\,\hat i\,).  
\eea   
Here   $\epsilon^{abc}$ are 
the completely antisymmetric structure constants. The left and the right electric fields or the angular momentum operators in (\ref{comrel1}) are related by a parallel transport as follows \cite{kogsus}: 
\begin{figure}
\begin{subfigure}{.5\textwidth}
\begin{center}
\begin{tikzpicture}[scale=1.2]
\coordinate (a) at (-2.,0);
\coordinate (b) at (2.5,0);
\draw[ultra thick, ->] (-2,0)--(0,0);
\draw[ultra thick] (a)--node[pos=.1,scale=1.6,gray] (){$\bullet$}node[pos=.1,scale=1.,above] (){$E_+(\vec{n};\, \hat{i}\,)$}node[pos=.5,scale=1.,above,sloped](e){$U_{\alpha\beta}( \vec n; \,\hat{i}\,)$}node[pos=.9,scale=1.6,gray] (){$\bullet$}node[pos=.9,scale=1.,above] (){$E_-(\vec{n}+\hat{i};\, \hat{i}\,)$}(b);
\node[below] at (a) {$\vec n$};
\node[below,scale=.9] at (2.5,0) {$\vec{n}+\hat{i}$};
\draw[opacity=.0] (0,-1.5)--(0,2);
\end{tikzpicture}
\end{center}
\caption{}
\end{subfigure}
\begin{subfigure}{.5\textwidth}
	\begin{center}
		\begin{tikzpicture}[scale=1.5]
		\draw[ultra thick](-1.5,0)--(-.75,0);
		\draw[ultra thick](-.75,0)--(.75,0);
		\draw[ultra thick](.75,0)--(1.5,0);
		\draw[ultra thick](0,-1.5)--(0,-.75);
		\draw[ultra thick](0,-.75)--(0,.75);
		\draw[ultra thick](0,.75)--(0,1.5);
		\node[]at (.2,.2){$\vec n$};
		\node[right]at (0,-1){$j(\vec n; \hat 4)$};
		\node[above]at (1,0){$j(\vec n; \hat 1)$};
		\node[above]at (-1,0){$j(\vec n; \hat 3)$};
		\node[right]at (0,1){$j(\vec n; \hat 2)$};
		\end{tikzpicture}
		\captionof{figure}{}\label{vertex}
	\end{center}
\end{subfigure}	
\caption{(a) Link holonomies $U(\vec n; \hat i))$ and  their conjugate electric fields $E_+(\vec n; \hat i)$ and $E_-(\vec n+\hat i; \hat i)$, (b) The SU(2) fluxes around a lattice site $\vec n$. }\label{link}
\end{figure}
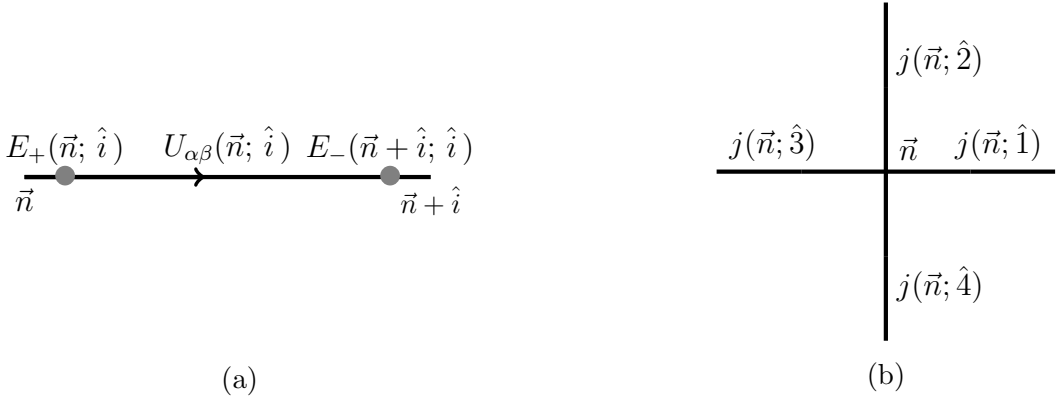
\bea 
E_-(\vec n+\hat i;\,\hat i\,) = -U^\dagger (\vec n;\,\hat i\,)\, E_+(\vec n;\,\hat i\,)\, U(\vec n;\,\hat i\,),
\label{lrrel} 
\eea 
where we have defined $ E_\pm \equiv \sum_{a} E^a_\pm \sigma^a$. It is easy to check that the left and right electric fields commute with each other and their magnitudes of  are equal 
\bea 
[E^a_+(n,\hat i),\, E^b_-(n+\hat i;\,\hat i \,)] =0,~~~~ 
\text{Tr} (E_+(\vec n;\,\hat i\,))^2 = \text{Tr} (E_-(\vec n+\hat i;\,\hat i\,))^2.
\label{ident12} 
\eea  
The identities (\ref{ident12}) imply that  
the complete set of commuting operators on a link $l=(\vec n, \hat i)$  in the 
dual electric fields representation 
 are 
$$\left[\vec E(l)^2, ~E_+^{a=3}(l), ~ E_-^{a=3}(l)\right].$$ 
In the above set  $\vec E^2(l)\equiv \vec E^2_+(\vec n, \hat i) = \vec E^2_-(\vec n+\hat i, \hat i)$. 
The corresponding eigenvectors, denoted by 
$\ket{j,m_+, m_-}_l \equiv \ket{j, m_+}_l \otimes \ket{j, m_-}_l$
\begin{align}
\begin{aligned}
 \vec E^2(l)~\ket{j,m_+, m_-}_l =\;j(l)\;\big(j(l)+1\big)~
 \ket{j,m_+, m_-}_l, \\
\phantom{lll}E_\pm^{a=3}(l) ~\ket{j,m_+, m_-}_l=\;  m_\pm (l)~\ket{j,m_+, m_-}_l. ~~~ 
 \label{eveq} 
\end{aligned}
\end{align}
The SU(2) gauge transformations at a lattice site $\vec n$ 
are \cite{kogsus}:  
\bea 
U(\vec n;\,\hat i\,) \rightarrow \Lambda(\vec n)~ U(\vec n;\,\hat i\,) ~\Lambda^\dagger(\vec n+\hat i\,); ~~ i=1,2. 
\label{gts} 
\eea 
In (\ref{gts}) $\Lambda(\vec n)$ are the SU(N) gauge degrees of freedom at lattice site $\vec n$.  The canonical commutation relations (\ref{comrel1}) imply that the generators of the above gauge transformations at site
$\vec n$ are: 
\begin{equation}\label{gtg}
{\cal G}^a(\vec n) \equiv \sum_{i=1}^{d}\left( E_+^{  a}(\vec n;\,\hat i\,)+E_-^{  a}(\vec n;\,\hat i\,) \right). 
\end{equation}
Physically, ${\cal G}^a(\vec n)$ represent the sum of 
all electric field or equivalently the angular momentum operators meeting at the lattice site $\vec n$. 
We also define the simplest gauge invariant  plaquette loop operators  
\vspace{-0.4cm} 
\bea \label{sunph}
\mathcal{U}_p=  U(\vec n;\,\hat i\,)U(\vec n+\hat i;\,\hat j\,)U^{\dagger}(\vec n+\hat j;\,\hat i\,)U^{\dagger}(\vec n;\,\hat j\,). 
\eea 
The Kogut Susskind $H_{KS}$ is 
\begin{equation}\label{sunksh}
H_{KS} = A \sum_{l} {\cal A}_{l} + B \sum_{p}{\cal B}_p,
\end{equation}
In (\ref{sunksh}),  $l$ and $p$ denote the 
links  and plaquettes, $A$ and $B$ are positive constants  
and   
\bea 
{\cal A}_l \equiv \sum_{a=1}^{3} {E}^a(l) {E}^a(l), ~~~~~ 
{\cal B}_p \equiv 1- \frac{1}{2} \text{Tr}~\mathcal{U}_p. 
\label{abes} 
\eea
Note that for SU(2) gauge group, ${\cal W}_p\equiv \frac{1}{2} \text{Tr}({\mathcal U}_p)$  is Hermitian and its eigenvalues are $\cos \omega_p$. The magnetic ordered states with ${\cal B}_p =0$ implies $\omega_p=0, 
\forall p$. 
\subsection{Wilson Loop and Spin Network States}
\label{spinn} 
We consider the Gauss law constraints at every lattice site $ \vec n$ 
\bea 
{\cal G}^a(\vec n)  \ket{\psi} =0, ~~~a=1,2,3.
\label{glc}  
\eea 
The gauge invariant states satisfying (\ref{glc}) are the physical states and belong to the physical Hilbert space ${\cal H}^p$.  On the other hand, the most general gauge invariant operators are 
the  Wilson loop operators ${\cal W}_{\cal C}$ 
around oriented  loops ${\cal C}$ which are 
 defined using SU(2) holonomies 
\begin{equation}
{\cal W}_{\cal C} = \frac{1}{2}\; \text{Tr} \left( \prod_{\vec l~\in\, {\cal C}} U(\,\vec l\,)\right). 
\label{wltc1} 
\end{equation}
Therefore, the most general gauge invariant (loop) states  can be written as 
\bea 
\ket{\;{\cal C}_1, {\cal C}_2,\cdots } \equiv {\cal W}_{{\cal C}_1} \; {\cal W}_{{\cal C}_2} \cdots ~ \ket{0}.   
\label{wls} 
\eea 
Here $\ket{0}$ is the gauge invariant strong coupling vacuum state defined by $\vec E^a_{\pm}(l) \ket{0}=0,~ \forall \; l$. The  loop states in (\ref{wls})  form a highly over-complete basis and satisfy the Mandelstam constraints \cite{mand, manuloop}. A complete and orthonormal basis set in ${\cal H}^p$ in SU(2) 
lattice gauge theory is the spin network basis. 
This  basis is best characterized in the dual 
formulation in terms of electric fluxes \cite{dual,manuloop} (see (\ref{eveq}) or angular momenta). 
We note that if there are $L$  SU(2) electric flux lines 
meeting at a vertex $\vec n$ then the Gauss law condition (\ref{glc}) states 
\begin{align}
\underbrace{\ket{j_1,j_2,j_{12}, \cdots, j_T =0,m_T=0}_{\vec n}}_{\equiv \ket{\;\vec J_{\vec n}}\;} =& \sum_{\text{all}~ m} C_{j_1,m_1;\,j_2,m_2}^{j_{12},\,m_{12}} \;C_{j_{12},m_{12};\,j_3,m_3}^{j_{123},\,m_{123}} \cdots  C_{j_{123\cdots L-1},m_{123\cdots L-1};\;j_L,\,m_L}^{j_T\equiv j_{12\cdots L} =0,\,m_T\equiv m_{12\cdots L}=0}  \nonumber \\
&~~~~~~~~~~~~~ \ket{j_1 m_1} \otimes \ket{j_2 m_2} \cdots \otimes\ket{j_L m_L} 
\label{cgc} 
\end{align} 
In (\ref{cgc}), $j_T \equiv j_{12\cdots L} =0, \; m_T \equiv m_{12\cdots L} =0$ and 
 $C_{j,\,m;\,\bar j,\,\bar m}^{J,M}$ are the 
standard Clebsch-Gordan coefficients.  
The Clebsch-Gordan couplings (\ref{cgc}) ensure that the spin network basis is gauge invariant ($j_T=0$) 
\bea 
{\cal G}^a(\vec n) |\vec J_{\vec n}\rangle=\underbrace{\sum_{i=1}^{d}\left( E_+^{  a}(\vec n;\,\hat i\,)+E_-^{  a}(\vec n;\,\hat i\,) \right)}_{\text{Total electric field at site} ~\vec n}|\vec J_{\vec n}\rangle=0, 
\label{gls}  
\eea 
and they 
form an orthonormal as well as  complete basis in ${\cal H}^{p}$, 
\bea 
\braket{ \vec K_{\vec n}}{ \vec J_{\vec n}}  = \delta_{\vec K_{\vec n}, \vec J_{\vec n}}, 
~~~ ~~~~~~\sum_{\vec J_{\vec n}}  \ket{ \vec J_{\vec n}}\, \bra{ \vec J_{\vec n}} = {\cal I}_{\,\vec n}.
\label{ocr}  
\eea 
In (\ref{ocr}) ${\cal I}_{\,\vec n}$ is an identity operator at lattice site $\vec n$. 
Therefore, on a finite dimensional lattice with ${\mathsf S}$ lattice sites, if we label all the 
vertices by  $\vec n \equiv v_1, v_2,\cdots , v_{{\mathsf S}}$, 
then 
 the most general state $\ket{\psi}$ satisfying (\ref{glc}) 
can be expanded in the spin network basis  
\bea 
|\psi\rangle  = \sum_{\{\vec J_{n_1},\, \vec J_{n_2},\, \cdots,\, \vec J_{n_{\mathsf S}} \}}^{} ~~\underbrace{\Phi \left(\vec J_{v_1}, \vec J_{v_2}, \cdots, \vec J_{v_{\mathsf S}} \right)}_{\text{amplitude}}  
~~\underbrace{\left\{
\prod_{i=1}^{{\mathsf S}} ~\otimes ~|\vec J_{v_i}\rangle
\right\}}_{\text{loop~ states}} \equiv \sum_{\{\vec J\}} \Phi (\vec J\,) ~\ket{\{\vec J\}\,}. 
\label{ls1tc1}
\eea  
In (\ref{ls1tc1}), the summation over $\{\vec J_v\}$ are constrained as  $j$ on each link is shared by its two  
vertices.  As an example in Figure \ref{vertex}, $j(\vec n ;\hat 1)\equiv j(\vec n+\hat1;\hat 3)$ and $j(\vec n ;\hat 2)\equiv j(\vec n+\hat2;\hat 4)$. 


\section{Wilson Loops and Wigner Identities}\label{wlwi} 

In this section we study   the gauge invariant 
eigenvalue equations  of  
$\mathcal{W}_{\cal C}$ in $d$ dimensions. These 
equations will be used later to derive various identities 
involving Wigner coefficients in $d=2,3$. 
The  state $\ket{\psi_\omega}$ satisfies   
\bea 
{\mathcal{W}}_{\cal C} ~\ket{\psi_\omega}= \cos \omega~ \ket{\psi_\omega}.   
\eea 
In the spin network basis we get    
\bea 
{\mathcal{W}}_{\cal C} ~\ket{\psi_\omega} = \sum_{\{\vec K\}}  \sum_{\{\vec J\}} ~\Phi_\omega (\vec J\;) ~{\cal M}^{\;({\cal C})}_{[\{\vec J\}\{\vec K\}]}~\ket{\{\vec K\}\,} = 
\cos \omega~ \sum_{\{\vec K\}}  \Phi_\omega (\vec K\;) ~\ket{\{\vec K\}\,}
\label{meq1} 
\eea 
In (\ref{meq1}) the  ${\cal M}^{~({\cal C})}_{[\{\vec J\}\;\{\vec K\}]}$ are the matrix elements of the magnetic flux operator  $\mathcal{W}_{{\cal C}}$
between the spin network states 
$\ket{\{\vec J\}\,}$ and $\ket{\{\vec K\}\,}$. Using the orthogonality properties of the spin network states
we get 
\bea \label{identity1}
\sum_{\{\vec K\}}\; {\cal M}^{~({\cal C})}_{[\{\vec J\}\{\vec K\}]}~ \Phi_\omega (\vec K\;) 
= \cos \omega ~\Phi_\omega  (\vec J\;)
\label{meve} 
\eea 
As we will see, the matrix elements ${\cal M}^{~({\cal C})}_{[\{\vec J\}~\{\vec K\}]}$ and the amplitudes $\Phi_\omega (\vec K\;)  $ are the  Wigner 3nj coefficients. Therefore,  the above identities represent the relationships  between different Wigner coefficients. Now we can iterate the above equation multiple times  to obtain a family of  identities among Wigner coefficients for a given loop ${\cal C}$ in a $d$ dimensional spin networks. From equation (\ref{identity1}) we get the most general identity
\bea \label{identity1aa}
\sum_{\{\vec K_1\},\{\vec K_2\},\, \cdots,\{\vec K_q\}} {\cal M}^{({\cal C})}_{[\{\vec J\}~\{\vec K_1\}]}\;{\cal M}^{({\cal C})}_{[\{\vec K_1\}~\{\vec K_2\}]}
\cdots  {\cal M}^{({\cal C})}_{[\{\vec K_{q-1}\}~\{\vec K_q\}]}\Phi_\omega  (\vec K_q\;)
= \; \left( \cos \omega \right)^q ~\Phi_\omega  (\vec J\;)
\eea 
In this work, we only consider the simple magnetic vacuum
cases ($\omega =0$) for small Wilson loops ${\cal C}$. 
In Appendix A, we have computed  the amplitudes $ \Phi\; (\vec J\;) $ for the 
magnetic vacuum states with $\omega =0$ using pure gauge conditions on lattices of different shapes, sizes. The matrix elements ${\cal M}_{[\{\vec J\},\; \{\vec K\}]}$ are computed in Appendix B.  
  The magnetic flux eigenvalue equation (\ref{meve}) then leads to non-trivial identities amongst Wigner coefficients. The more general identities  can be similarly derived.   In the following  
sections, we construct  simple models on finite lattices 
to  implement the above ideas.
\subsection{A Toy Model on a Tetrahedron}\label{tetra} 
 We know that the simplest   3nj Wigner coefficients are the $6j$ coefficients. The six values of $j$ can represent SU(2) flux values on the six edges of a tetrahedron. Therefore, we consider a tetrahedron ${\cal T}$ (see Figure \ref{tetra6j}) with vertices $v=a, b, c, d$ and  oriented triangular plaquettes\footnote{By oriented triangle $\triangle_{abc}$ we mean that it is traversed in the direction $a \rightarrow b \rightarrow c \rightarrow a$} or the smallest Wilson loops  ${\cal C} \equiv p$ = $\triangle_{abc},\; \triangle_{abd},\; \triangle_{adc},\; \triangle_{bcd}$. We now analyze the Kogut Susskind  Hamiltonian  on ${\cal T}$
\bea 
H= A \sum_{l} {\cal A}_l +B \sum_{p} {\cal B}_{p}.
\label{tcht} 
\eea  
Above A and B are $+ve$ constants and  
\begin{align}
\begin{aligned}
{\cal A}_l  =\; &\sum_{{\mathsf a}=1}^{3} {E}^a(l) \;{E}^{\mathsf a}(l),~~ ~~~~~~~~~~~~~\;l=ab, ac, ad, bc, bd, cd; \\
{\cal B}_p  = \;&  \;\Big(1- \frac{1}{2}\; {\text Tr}\;\triangle_p\Big), ~~~~~~~~~~~~ \triangle_p= \triangle_{{abc}},\; \triangle_{{bad}}, \;\triangle_{{bdc}}, \; \triangle_{{dac}}.
\end{aligned}
\end{align}
The magnetic field terms are
\bea \label{ab=0}
\triangle_{abc} = U_{ab}\; U_{bc} \; U_{ca}, ~~~~~~~~~~ \triangle_{adb} = U_{ad}\; U_{db} \; U_{ba} \nonumber \\
\triangle_{acd} = U_{ac}\; U_{cd} \; U_{da}, ~~~~~~~~~~ \triangle_{bdc} = U_{bd}\; U_{dc} \; U_{ca}
\eea
All the link operators or the link holonomies along the 6 edges  
of ${\cal T}$ are chosen in the fundamental $j\!=\!\frac{1}{2}$ representation. 
The SU(2) transformations at the 
at the 4 vertices are  are generated by the 4 Gauss law generators
\begin{align}
\begin{aligned}
{\cal G}^{ \mathsf{a}}(a) \equiv & \, E^{ \mathsf{a}}(a,1)+E^{\mathsf{a}}(a,2) + E^{ \mathsf{a}}(a,3),~~~~ {\cal G}^\mathsf{a}(b) \equiv  E^{\mathsf{a}}(b,3)+E^{\mathsf{a}}(b,4) + E^{\mathsf{a}}(b,5), \\ 
  {\cal G}^\mathsf{a}(c) \equiv &\, E^{\mathsf{a}}(c,2)+E^{\mathsf{a}}(c,4) + E^{\mathsf{a}}(c,6),  
~~~~\; {\cal G}^\mathsf{a}(d) \equiv  E^{\mathsf{a}}(d,1)+E^{\mathsf{a}}(d,5) + E^{\mathsf{a}}(d,6).
\end{aligned}
\end{align}
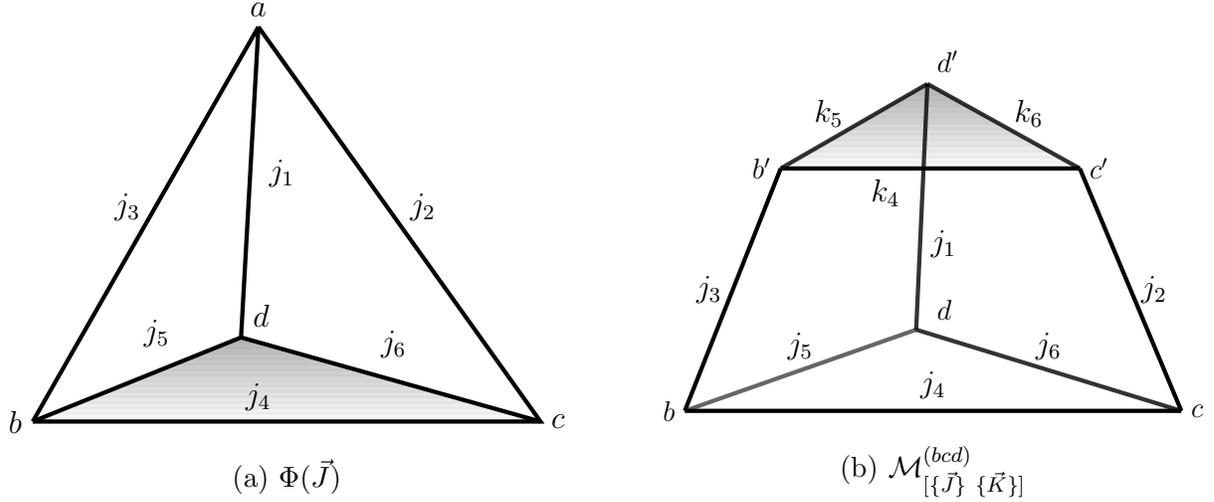
\begin{figure}[t]
\begin{subfigure}[b]{.5\textwidth}
	\begin{center}
	\begin{tikzpicture}[scale=1.5]
\coordinate (b) at (-2,0,0);
\coordinate (c) at (2.5,0,0);
\coordinate (d) at (0,.9,.4);
\coordinate (a) at (0,3.5,0);
\shade[fill={gray},top color=gray!60,bottom color=gray!02] (b)--(c)--(d);
\draw[ultra thick] (a)--(c)--(b);
\draw[ultra thick] (b)--(a);
\draw[ultra thick] (a)--(d)--(b);
\draw[ultra thick] (c)--(d);
\node[above] at (a) {$a$};
\node[left] at (b) {$b$};
\node[right] at (c) {$c$};
\node[above] at (0.18,.9,.4) {$d$};
\node[left] at (-.95,1.9) {$j_3$};
\node[below] at (-.9,1.) {$j_5$};
\node[above] at (0,0.) {$j_4$};
\node[right] at (0.,2.2) {$j_1$};
\node[right] at (1.25,1.9) {$j_2$};
\node[below] at (1.2,.9) {$j_6$};
	\end{tikzpicture}
\caption{ $\Phi(\vec J)$}\label{tetra6j}
\end{center}
	\end{subfigure}
	\begin{subfigure}[b]{.5\textwidth}
	\begin{center}
	\begin{tikzpicture}[scale=1.5]
\coordinate (a') at (-1.3,.45);
\coordinate (b') at (1.35,.45);
\coordinate (c') at (0,1.2);
\coordinate (a) at (-2.15,-1.7);
\coordinate (b) at (2.25,-1.7);
\coordinate (c) at (-0.1,-.98);
\coordinate (d) at (0,2.5);
\coordinate (o) at (.4,1.);
\shade[fill={gray},top color=gray!60,bottom color=gray!02] (a')--(c')--(b');
\draw[ultra thick] (a)--node[pos=.5,scale=1.,above] (){$j_4$}(b);
\draw[ultra thick,opacity=.6] (a)--node[pos=.5] (e){}(c);
\node[scale=1.,above] at (e){$j_5$};
\draw[ultra thick,opacity=.8] (b)--node[pos=.5] (f){}(c);
\node[scale=1.,above] at (f){$j_6$};
\draw[ultra thick] (a)--node[pos=.5] (g){}(a');
\node[scale=1.,left] at (g){$j_3$};
\draw[ultra thick] (b)--node[pos=.5] (h){}(b');
\node[scale=1.,right] at (h){$j_2$};
\draw[ultra thick,opacity=.8] (c)--node[pos=.35] (i){}(c');
\node[scale=1.,right] at (i){$j_1$};
\draw[ultra thick] (a')--node[pos=.35] (j){}(b');
\node[scale=1.,right,below] at (j){$k_4$};
\draw[ultra thick,opacity=.8] (a')--node[pos=.5,above] (k){}(c');
\node[scale=1.,left] at (k){$k_5$};
\draw[ultra thick,opacity=.8] (b')--node[pos=.5,above] (l){}(c');
\node[scale=1.,right] at (l){$k_6$};
\node[scale=.9,left] at (a) {$b$};
\node[scale=.9,right] at (b) {$c$};
\node[scale=.9,right] at (.,-.8) {$d$};
\node[scale=.9,left] at (a') {$b'$};
\node[scale=.9,right] at (b') {$c'$};
\node[scale=.9,right] at (.,1.4) {$d^\prime$};
\end{tikzpicture}
\caption{$ {\cal M}^{(bcd)}_{[\{\vec J\}~\{\vec K\}]}$}\label{tetra9j}
\end{center}
	\end{subfigure}
\caption{Gauge theory on a tetrahedron: (a) The tetrahedron  with the j labels  represents the Wigner 6j coefficients which are the magnetic ground state   amplitudes $\Phi(\vec J)$ in  (\ref{abn1}) in the spin network basis,  (b) Matrix elements ${\cal M}^{(bcd)}_{[\{\vec J\}~\{\vec K\}]}$ in (\ref{xvb2}).}
\end{figure}
We will work with magnetically   ordered  
gauge invariant states  $\ket{\psi_0}$ with  0 magnetic fields. They satisfy 
\bea 
{\cal G}^a(v)~\ket{\psi_{\omega=0}} =0, ~~\forall \; v,~~~~~~~~~~ {\cal B}_p~ \ket{\psi_{\omega=0}} =0, ~~\forall \; p. 
\label{gscs} 
\eea 
The state ${\psi_0}$  can be expanded in the spin network\footnote{In this simple case, the spin network states at the vertex $a$ are
	\begin{equation}
	\ket{j_1,j_2,j_3}_a=(-1)^{j_1-j_2+j_3}\sum_{m_1,\,m_2,\,m_3}  \left( \begin{array}{ccc}
	\!\!j_1&j_2& j_3\!\!\\
	\!\!m_1&m_2& m_3\!\! 
	\end{array} \right)\ket{j_1,m_1}\otimes\ket{j_2,m_2}\otimes\ket{j_3,m_3}.
	\end{equation} Similarly we can construct these states at the other 3 vertices.}  basis  (\ref{cgc})
\bea
\ket{\psi_0} & = & \sum_{j_1,\,j_2,\,\cdots,\, j_6}\underbrace{ {\Phi}\left(j_1,j_2,\cdots,j_6\right)}_{\text{amplitude on }~{\cal T}} ~ 
\underbrace{\ket{j_1\;j_2\;j_3}_a \otimes\ket{j_3\;j_4\;j_5}_b \otimes \ket{j_2\;j_4\;j_6}_c \otimes \ket{j_1\;j_5\;j_6}_d}_{\text{spin network} ~ \text{on} ~{\cal T}}  \nonumber \\
&\equiv &  ~~\sum_{\{\vec J\}} \; \Phi (\vec J\;)~ \ket{\{\vec J\}}
\eea 
 The amplitudes $\Phi({\vec J})$, which ensure the magnetic fields 
 are zero, can be constructed by choosing pure gauge conditions on every link. The detail of this calculations are given in Appendix A. The final result is 
\bea \label{tetraamplitude}
\Phi(\vec J\;) \equiv \Phi\;\left(j_1,j_2,\cdots,j_6\right) = \Pi(j_1,j_2,\cdots ,j_6) ~
\left\{ \begin{array}{cccc}
\!\!{j}_{1} & \!{j}_{2} \!&\! j_3 \!\!\!\\
\!\!{j}_{4} &\! j_{5} \!&\! j_6\!\!\!\\
\end{array} \right \}.
\label{abn1}
\eea 
In (\ref{abn1}), $\Pi(a,b,\cdots) \equiv \sqrt{(2a+1)(2b+1)\cdots}$.  The $6j$ coefficients or the amplitude 
 in (\ref{abn1})  is  shown in Figure-\ref{tetra6j}.
Using the generalized Wigner Eckart theorem, the matrix  elements $ {\cal M}^{(bcd)}_{[\{\vec J\}\{\vec K\}]}$ can be computed (see Appendix B).
\bea 
{\cal M}^{(bcd)}_{[\{\vec J\}~\{\vec K\}]} & = &~~~ \frac{1}{2}\;\bra{\{\vec J\}} \;{\text Tr\;}\triangle_{bcd}\; \ket{\{\vec K\}} \nonumber \\ 
& = & 
{M}^{(bcd)}(\vec J, \vec K) ~\underbrace{\left[\begin{array}{cccccc}
\!k_4\!\! & & \!\!{k}_5 \!\!&  & \!\!k_6\!\!&\!\!\\
&\!\!j_3\!\!& &\!\!j_1\!\!& &\!\! j_2\!\!\!\\  
\!{j}_{4}\!\! & &\!\! {j}_{5}\!\! &  & \!\!j_6 \!\!& \!\!\\
\end{array} \right]}_{9j ~\text{ coefficient~ of }~2^{\text{nd}}~\text{kind}} ~\prod_{l=4,5,6}  
\left\{j_l,\;k_l,\;\frac{1}{2}\right\}. 
\label{xvb2} 
\eea 
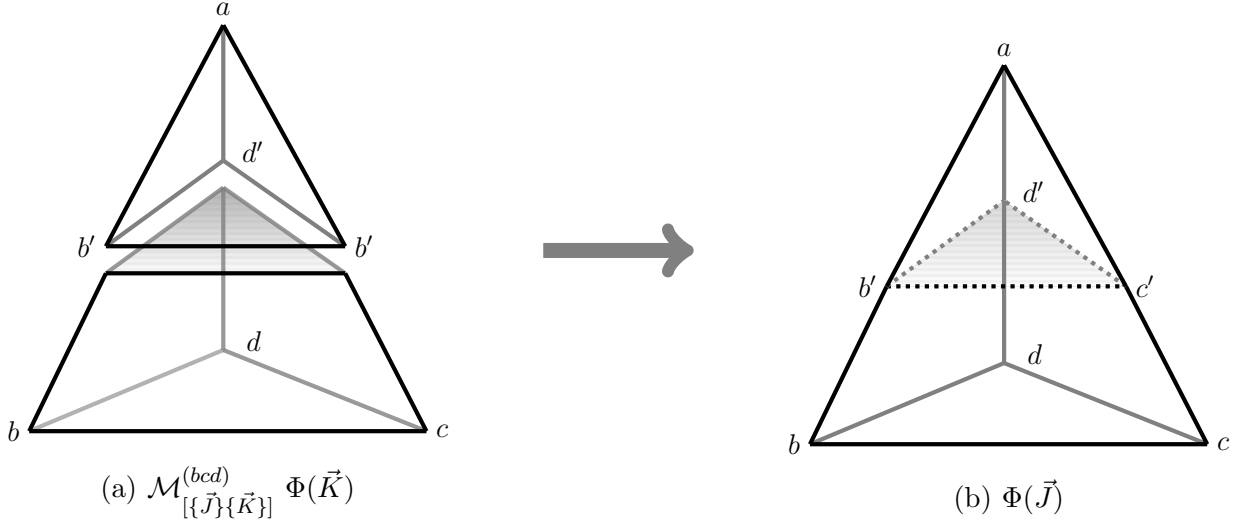
\begin{figure}
\begin{subfigure}[b]{.4\textwidth}
\begin{center}
\begin{tikzpicture}[scale=1.2]
\coordinate (a') at (-1.3,.05);
\coordinate (b') at (1.35,.05);
\coordinate (c') at (0,1.);
\coordinate (a) at (-2.15,-1.7);
\coordinate (b) at (2.25,-1.7);
\coordinate (c) at (0.,-.8);
\coordinate (d) at (0,2.5);
\coordinate (o) at (.4,1.);
\shade[fill={gray},top color=gray!60,bottom color=gray!02] (a')--(c')--(b');
\draw[ultra thick,gray,opacity=.6] (a)--(c);
\draw[ultra thick,gray,opacity=.8] (b)--(c);
\draw[ultra thick,gray,opacity=.8] (c)--(c');
\draw[ultra thick,gray,opacity=.8] (a')--(c');
\draw[ultra thick,gray,opacity=.8] (b')--(c');
\draw[ultra thick] (a)--(b);
\draw[ultra thick] (a)--(a');
\draw[ultra thick] (b)--(b');
\draw[ultra thick] (a')--(b');
\node[scale=.9,left] at (a) {$b$};
\node[scale=.9,right] at (b) {$c$};
\node[scale=.9,right] at (.15,-.7) {$d$};
\begin{scope}[shift={(0,.3)}]
\coordinate (a') at (-1.3,.05);
\coordinate (b') at (1.35,.05);
\coordinate (c') at (0,1.);
\coordinate (a) at (-2.15,-1.7);
\coordinate (b) at (2.25,-1.7);
\coordinate (c) at (0.,-.8);
\coordinate (d) at (0,2.5);
\coordinate (o) at (.4,1.);
\draw[ultra thick,gray] (a')--(c');
\draw[ultra thick,gray] (b')--(c');
\draw[ultra thick,gray] (c')--(d);
\draw[ultra thick] (a')--(b');
\draw[ultra thick] (a')--(d);
\draw[ultra thick] (b')--(d);
\node[scale=.9,above] at (d) {$a$};
\node[scale=.9,left] at (a') {$b'$};
\node[scale=.9,right] at (b') {$b'$};
\node[scale=.9,right] at (0.1,1.1) {$d'$};
\end{scope}
	\end{tikzpicture}
\caption{${\cal M}^{(bcd)}_{[\{\vec J\}\{\vec K\}]}~\Phi(\vec K)$}
\end{center}
	\end{subfigure}
\begin{subfigure}[b]{.2\textwidth}
\begin{center}
\begin{tikzpicture}
\draw[line width=2.mm,->,gray] (-1,.0)--(1,.0);
\end{tikzpicture}
\end{center}
\vspace{3cm}
\end{subfigure}	
\begin{subfigure}[b]{.4\textwidth}
	\begin{center}
	\begin{tikzpicture}[scale=1.2]
\coordinate (a') at (-1.3,.05);
\coordinate (b') at (1.35,.05);
\coordinate (c') at (0,1.);
\coordinate (a) at (-2.15,-1.7);
\coordinate (b) at (2.25,-1.7);
\coordinate (c) at (0.,-.8);
\coordinate (d) at (0,2.5);
\coordinate (o) at (.4,1.);
\shade[fill={gray},top color=gray!30,bottom color=gray!02] (a')--(c')--(b');
\draw[ultra thick,gray] (a)--(c);
\draw[ultra thick,gray] (b)--(c);
\draw[ultra thick,gray] (c)--(c');
\draw[ultra thick,gray] (c')--(d);
\draw[ultra thick] (a)--(b);
\draw[ultra thick] (a)--(a');
\draw[ultra thick] (b)--(b');
\draw[ultra thick,dotted] (a')--(b');
\draw[ultra thick,gray,dotted] (a')--(c');
\draw[ultra thick,gray,dotted] (b')--(c');
\draw[ultra thick] (a')--(d);
\draw[ultra thick] (b')--(d);
\node[scale=.9,left] at (a) {$b$};
\node[scale=.9,right] at (b) {$c$};
\node[scale=.9,right] at (.15,-.7) {$d$};
\node[scale=.9,above] at (d) {$a$};
\node[scale=.9,left] at (a') {$b'$};
\node[scale=.9,right] at (b') {$c'$};
\node[scale=.9,right] at (0.1,1.1) {$d'$};
	\end{tikzpicture}
\caption{$\Phi(\vec J)$}
\end{center}
\end{subfigure}
\caption{Graphical representation of the Wilson loop eigenvalue equation (\ref{meve}) on a tetrahedron: ${\cal M}^{(bcd)}_{[\{\vec J\}\{\vec K\}]}~\Phi(\vec K)= \Phi(\vec J)$ leading to Wigner coefficient identities (\ref{boxabn1}): \!\!\! 9j $\times$ 6j=6j in (\ref{fi6j}).}\label{tetrafusion}
\end{figure}
In (\ref{xvb2}),  $${M}^{(bcd)}(\vec J, \vec K) = 
\delta_{j_1,k_1} \;\delta_{j_2,k_2}\; \delta_{j_3,k_3} 
~\frac{\Pi\left(j_4,j_5,j_6,k_4,k_5,k_6\right)}{\Pi^4(j=\frac{1}{2})}.$$ and 
\bea 
{\left\{a,\;b,\;c\right\}}  =\begin{cases}1& ~~{\text if}~  (a,\;b,\;c)~ {\text{\;form\; a \;triangle}},\\
0&~~~~~~~~~{\text{otherwise}}.
\end{cases}  
\label{tcon}
\eea 
Thus the  matrix elements of the triangular plaquette operator 
are 9j Wigner coefficients of the second kind \cite{varsha,yutsis}.
\subsubsection{6j-9j and 6j-12j Wigner Identities}
We now substitute the amplitude (\ref{abn1}) and the matrix elements (\ref{xvb2}) in the equation (\ref{meq1}) 
to get the 
first identity derived using the  simplest Wilson loop operator ${\cal W}_{bcd} =\frac{1}{2} {\text Tr}\triangle_{bcd}$ in the 
fundamental $j=\frac{1}{2}$ representation  
\bea 
\label{itype11}
\sum_{\{k\}'} 
{\Pi^2(k_4,k_5,k_6)}
\left[\begin{array}{cccccc}
\!k_4\!\! & & \!\!{k}_5 \!\!&  & \!\!k_6\!\!&\!\!\\
&\!\!j_3\!\!& &\!\!j_1\!\!& &\!\! j_2\!\!\!\\  
\!{j}_{4}\!\! & &\!\! {j}_{5}\!\! &  & \!\!j_6 \!\!& \!\!\\
\end{array} \right]\left\{ \begin{array}{cccc}
\!\!{j}_{1} & \!{j}_{2} \!&\! j_3 \!\!\!\\
\!\!{k}_{4} &\! k_{5} \!&\! k_6\!\!\!\\
\end{array} \right \} =  \Pi^4 ({\frac{1}{2}}) 
\left\{ \begin{array}{cccc}
\!\!{j}_{1} & \!{j}_{2} \!&\! j_3 \!\!\!\\
\!\!{j}_{4} &\! j_{5} \!&\! j_6\!\!\!\\
\end{array} \right \}
\label{fi6j} 
\eea 
The summation over $\{k\}'$ means 
that the values of $(k_4,k_5,k_6)$ are restricted by the triangular constraints 
 $\{j_4,\;k_4,\;\frac{1}{2}\}, ~  \{j_5,\;k_5,\;\frac{1}{2}\}, ~\{j_6,\;k_6,\;\frac{1}{2}\}$ given in (\ref{tcon}).  In other words, $k_4 = j_4\pm\frac{1}{2},k_5 = j_5\pm\frac{1}{2},k_6 
=j_6\pm\frac{1}{2}$.  The reason for the appearance of the factor $\Pi^4(\frac{1}{2})$ in (\ref{fi6j}) is that the SU(2) flux raising and lowering 
Wilson loop operator ${\cal W}_{bcd}$ 
was chosen to be in the 
fundamental $j=\frac{1}{2}$ spin representation. If we use the plaquette  operator in an arbitrary spin $s$ representation, ${\cal W}^{(s)}_{bcd}=  \frac{1}{(2s+1)}{\text Tr} \left(\triangle^{(s)}_{bcd}\right)$, then we get 
\bea \label{itype11p}
\sum_{\{k\}'} 
\Pi^2(k_4,k_5,k_6) 
 \left[\begin{array}{cccccc}
\!k_4\!\! & & \!\!{k}_5 \!\!&  & \!\!k_6\!\!&\!\!\\
&\!\!j_3\!\!& &\!\!j_1\!\!& &\!\! j_2\!\!\!\\  
\!{j}_{4}\!\! & &\!\! {j}_{5}\!\! &  & \!\!j_6 \!\!& \!\!\\
\end{array} \right]\left\{ \begin{array}{cccc}
\!\!{j}_{1} & \!{j}_{2} \!&\! j_3 \!\!\!\\
\!\!{k}_{4} &\! k_{5} \!&\! k_6\!\!\!\\
\end{array} \right \}=  {\Pi^4(s)}
\left\{ \begin{array}{cccc}
\!\!{j}_{1} & \!{j}_{2} \!&\! j_3 \!\!\!\\
\!\!{j}_{4} &\! j_{5} \!&\! j_6\!\!\!\\
\end{array} \right \}
\label{fi6j1} 
\eea 
In (\ref{fi6j1})  the sums 
are restricted by the  triangular 
constraints: $\{j_4,\;k_4,\;s\},  \{j_5,\;k_5,\;s\}, ~\{j_6,\;k_6,\;s\}.$ 
Theses identities can also be represented geometrically 
as a fusion of tetrahedron and  tetrahedron frustum into a 
tetrahedron as shown in  Figure-\ref{tetrafusion}. In Appendix \ref{proof}, we prove (\ref{fi6j1}) explicitly with the help of 
known results given in the book by Varshalovich et. al. \cite{varsha}. 

We can also obtain identities involving higher Wigner coefficients by applying larger Wilson loops. 
As an example, we consider 2 plaquette Wilson loop $\mathcal{W}_{\cal C} \equiv \mathcal{W}_{bcda}= \frac{1}{2}{\text Tr}\left(\triangle_{bcd}\triangle_{bda}\right)$ on $\ket{\psi_0}$.  As this loop operator changes 4 edges, we get $12j$ Wigner coefficients of second kind \cite{varsha,yutsis} as the matrix elements  
\bea 
{\cal M}^{(bcda)}_{[\{\vec J\}\{\vec K\}]} & = & \frac{1}{2} \; \bra{\{\vec J\}} \;{\text Tr} \left(\triangle_{bcda}\right)\; \ket{\{\vec K\,\}} \nonumber \\ 
& = & 
{M}^{(bcda)}(\vec J, \vec K) \underbrace{\left[ \begin{array}{cccccccc}
\!{j}_{1}\!\! &   & \!\!{j}_{3}\!\! &  &\!\! j_4\!\! & &\!\!j_6\!\!&\\
& \!\!j_2\!\! & & \!\!j_5  \!\!& & \!\!j_2&\!\! &\!\!j_5\!\!\!\\  
\!{k}_{1}\!\! & & \!\!{k}_{3}\!\! &  & \!\!k_4\!\! &&\!\!k_6\!\!&\!\!\\
\end{array} \right]}_{12j~\text{coefficient of} ~2^{\text{nd}}~  \text{kind}} ~\prod_{l=1,3,4,6}  \left\{j_l,\;k_l,\;\frac{1}{2}\right\}.~~~~~ 
\label{xvb1} 
\eea 
In (\ref{xvb1}),  $${M}^{(bcda)}(\vec J, \vec K) = 
\delta_{j_2,\,k_2} \;\delta_{j_5,\,k_5}~ 
\frac{\Pi\left(j_1,\; j_3,\; j_4,j_6,\; k_1,\; k_3,\; k_4,\; k_6\right)}{\Pi^4(\frac{1}{2})}$$ and the matrix elements of the triangular plaquette operator 
is $12j$ coefficient of second kind.
We thus get the identity involving Wigner $12j$ and 
Wigner $6j$ coefficients:
\bea 
\sum_{\{k\}'} 
\Pi^2(k_1,k_3,k_4,k_6)  
\left[ \begin{array}{cccccccc}
\!{j}_{1}\!\! &   & \!\!{j}_{3}\!\! &  &\!\! j_4\!\! & &\!\!j_6\!\!&\\
& \!\!j_2\!\! & & \!\!j_5  \!\!& & \!\!j_2&\!\! &\!\!j_5\!\!\!\\  
\!{k}_{1}\!\! & & \!\!{k}_{3}\!\! &  & \!\!k_4\!\! &&\!\!k_6\!\!&\!\!\\
\end{array} \right]\left\{ \begin{array}{cccc}
\!\!{k}_{1} & \!{j}_{2} \!&\! k_3 \!\!\!\\
\!\!{k}_{4} &\! j_{5} \!&\! k_6\!\!\!\\
\end{array} \right \} = \Pi^4 ({\frac{1}{2}}) 
\left\{ \begin{array}{cccc}
\!\!{j}_{1} & \!{j}_{2} \!&\! j_3 \!\!\!\\
\!\!{j}_{4} &\! j_{5} \!&\! j_6\!\!\!\\
\end{array} \right \}
\label{6j12j} 
\eea 
Here the 4 summations in $\{k\}'$  are restricted over the triangular constraints defined in (\ref{tcon}): 
$\{j_1,\;k_1,\;\tfrac{1}{2}\},  \{j_3,\;k_3,\;\frac{1}{2}\}, \{j_4\;k_4,\;\frac{1}{2}\}, ~ \{j_6,\;k_6,\;\frac{1}{2}\}.$
\label{trcon} 
We can further generalize (\ref{6j12j}) for a general plaquette
in an arbitrary spin $s$ representation $\mathcal{W}^{(s)}_{\cal C} \equiv \mathcal{W}^{(s)}_{bcda}= \frac{1}{2}\; {\text Tr}\triangle^{(s)}_{bcd}\; \triangle^{(s)}_{bda}$
by simply replacing 
$(\frac{1}{2})$ by $(s)$ in (\ref{6j12j}) as well as in the above triangular constraints  as was done in the previous $(6j-9j)$ case.  
\subsection{A model on a cube}\label{box}
We now consider a cube  with twelve angular momentum $j_1,\;j_2,\dots, j_{12}$ assigned to its 12 edges as  shown in Figure \ref{boxi}.
The state $\ket{\psi_0}$  can be expanded in the spin network basis  
\begin{align}
\ket{\psi_0}  = & \sum_{j_1,j_2,\cdots j_6}\underbrace{ {\Phi}\left(j_1,j_2,\cdots,j_6\right)}_{\text{amplitude on cube}} ~ \ket{j_1\;j_2\;j_5}_a \otimes\ket{j_2\;j_3\;j_6}_b \otimes \ket{j_3\;j_4\;j_7}_c \otimes \ket{j_4\;j_1\;j_8}_d\\
&\qquad\qquad\qquad\qquad\qquad\;\;\underbrace{\ket{j_5\;j_9\;j_{10}}_e \otimes\ket{j_6\;j_{10}\;j_{11}}_f \otimes \ket{j_7\;j_{11}\;j_{12}}_g \otimes \ket{j_8\;j_{9}\;j_{12}}_h}_{\text{spin network} ~\ket{\vec J\,}~ \text{on} ~ \text{cube}}  \nonumber \\
\equiv &  ~~\sum_{\{\vec J\}} \; \Phi (\vec J\;)~ \ket{\{\vec J\} \,}
\end{align}
The amplitude $\Phi \left(j_1,j_2,\cdots,j_6\right)$ are now 
 given in terms of the Wigner 12j coefficients 
\bea \label{boxamplitude}
\Phi (\vec J\;) \equiv \Phi\;\left(j_1,j_2,\cdots,j_{12}\right) = \Pi(j_1,j_2,\cdots ,j_{12}) ~
\left[ \begin{array}{ccccccccc}
\!{j}_{1} \!\!&&\!\! {j}_{2}\!\! &&\!\! j_3\!\!&&\!\!j_4\!\! &\!\!\\
 &\!\!j_5\!\!&& \!\!j_{6} \!\!&& \!\!j_7\!\!&&\!\!j_8\!\!\\
\!{j}_{9}\!\! &&\!\! j_{10}\!\! &&\!\! j_{11}\!\!&&\!\!j_{12}\!\!&\\
\end{array} \right]
\label{abn2}
\eea 
\begin{figure}
\begin{center}
\begin{tikzpicture}[scale=1.6]
\coordinate (a) at (0,0,0);
\coordinate (b) at (2.5,0,0);
\coordinate (c) at (2.5,0,-3);
\coordinate (d) at (0,0,-3);
\coordinate (e) at (0,2.5,0);
\coordinate (f) at (2.5,2.5,0);
\coordinate (g) at (2.5,2.5,-3);
\coordinate (h) at (0,2.5,-3);
\shade[fill={gray},top color=gray!40,bottom color=gray!02] (a)--(b)--(c)--(d);
\draw[ultra thick] (a)--node[pos=.5,scale=1.,above,sloped]{$j_1$}(b);
\draw[ultra thick] (b)--node[pos=.5,scale=1.,right]{$j_2$}(c);
\draw[ultra thick] (c)--node[pos=.7,scale=1.,above,sloped]{$j_3$}(d);
\draw[ultra thick] (d)--node[pos=.5,scale=1.,above]{$j_4$}(a);
\draw[ultra thick] (e)--node[pos=.7,scale=1.,above,sloped]{$j_9$}(f);
\draw[ultra thick] (f)--node[pos=.5,scale=1.,left]{$j_{10}$}(g);
\draw[ultra thick] (g)--node[pos=.5,scale=1.,above,sloped]{$j_{11}$}(h);
\draw[ultra thick] (h)--node[pos=.5,scale=1.,left]{$j_{12}$}(e);
\draw[ultra thick] (a)--node[pos=.5,scale=1.,left]{$j_{8}$}(e);
\draw[ultra thick] (b)--node[pos=.65,scale=1.,right]{$j_{5}$}(f);
\draw[ultra thick] (c)--node[pos=.4,scale=1.,right]{$j_{6}$}(g);
\draw[ultra thick] (d)--node[pos=.4,scale=1.,right]{$j_{7}$}(h);
\node[left] at (a) {$d$};
\node[right] at (2.6,0,0) {$a$};
\node[right] at (c) {$b$};
\node[left] at (-.1,0,-3) {$c$};
\node[left] at (e) {$h$};
\node[right] at (2.6,2.5,0) {$e$};
\node[above] at (g) {$f$};
\node[above] at (h) {$g$};
	\end{tikzpicture}
\caption{Gauge theory on a box. The box with the j labels   represents the Wigner 12j coefficients which are the magnetic ground state   amplitudes (\ref{abn2}) in the spin network basis.}\label{boxi}
\end{center}
\caption{}
\end{figure}
\subsubsection{12j-12j and 12j-18j Wigner Identities}
We consider the simplest Wilson plaquette loop operator in $j\!=\!\frac{1}{2}$ representation:   
${\cal W}_{abcd} = \frac{1}{2} \text{Tr}~\square_{abcd}$. 
The matrix elements $ {\cal M}^{(abcd)}_{[\{\vec J\}~\{\vec K\}\,]}$ are
\bea 
{\cal M}^{(abcd)}_{[\{\vec J\}~\{\vec K\}\,]} & = & \frac{1}{2}\bra{\{\vec J\}\,} \;\text{Tr}\;\square_{abcd}\; \ket{\{\vec K\}\,} \nonumber \\ 
& = & 
{M}^{(abcd)}(\vec J, \vec K) ~\underbrace{\left[ \begin{array}{ccccccccc}
\!{j}_{1} \!\!&&\!\! {j}_{2}\!\! &&\!\! j_3\!\!&&\!\!j_4\!\! &\!\!\\
 &\!\!j_5\!\!&& \!\!j_{6} \!\!&& \!\!j_7\!\!&&\!\!j_8\!\!\\
\!{k}_{1}\!\! &&\!\! k_{2}\!\! &&\!\! k_{3}\!\!&&\!\!k_{4}\!\!&\\
\end{array} \right]}_{12j ~\text{ coefficient~ of }~2^{\text{nd}}~\text{kind}} ~\prod_{l=1}^4 \{j_l,k_l,\frac{1}{2}\} 
\label{boxxvb} 
\eea 
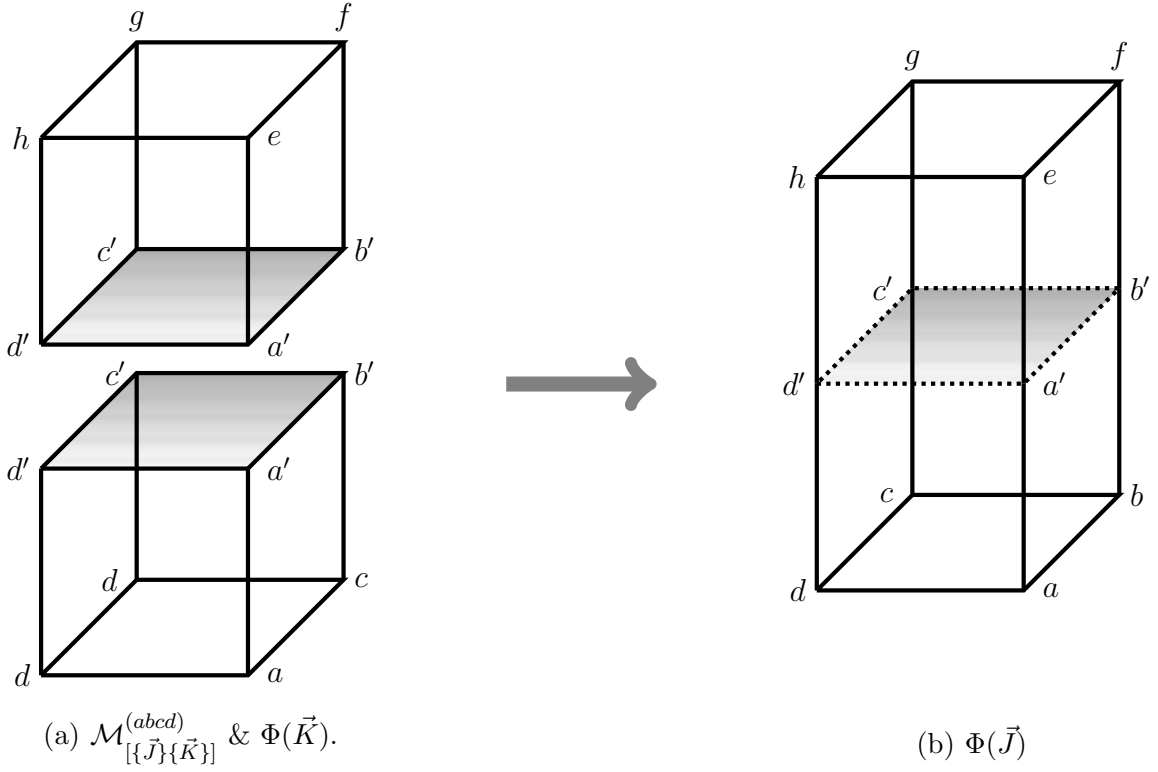
\begin{figure}[t]
\begin{subfigure}[b]{.4\textwidth}
\begin{center}
\begin{tikzpicture}[scale=1.1]
\coordinate (a) at (0,0,0);
\coordinate (b) at (2.5,0,0);
\coordinate (c) at (2.5,0,-3);
\coordinate (d) at (0,0,-3);
\coordinate (e) at (0,2.5,0);
\coordinate (f) at (2.5,2.5,0);
\coordinate (g) at (2.5,2.5,-3);
\coordinate (h) at (0,2.5,-3);
\shade[fill={gray},top color=gray!60,bottom color=gray!02] (a)--(b)--(c)--(d);
\draw[ultra thick](a)--(b)--(c)--(d)--(a);
\draw[ultra thick](e)--(f)--(g)--(h)--(e);
\draw[ultra thick](a)--(e);
\draw[ultra thick](b)--(f);
\draw[ultra thick](c)--(g);
\draw[ultra thick](d)--(h);
\node[left] at (a) {$d'$};
\node[right] at (2.6,0,0) {$a'$};
\node[right] at (c) {$b'$};
\node[left] at (-.1,0,-3) {$c'$};
\node[left] at (e) {$h$};
\node[right] at (2.6,2.5,0) {$e$};
\node[above] at (g) {$f$};
\node[above] at (h) {$g$};
\begin{scope}[shift={(0,-4)}]
\coordinate (a) at (0,0,0);
\coordinate (b) at (2.5,0,0);
\coordinate (c) at (2.5,0,-3);
\coordinate (d) at (0,0,-3);
\coordinate (e) at (0,2.5,0);
\coordinate (f) at (2.5,2.5,0);
\coordinate (g) at (2.5,2.5,-3);
\coordinate (h) at (0,2.5,-3);
\shade[fill={gray},top color=gray!60,bottom color=gray!02] (e)--(f)--(g)--(h);
\draw[ultra thick](a)--(b)--(c)--(d)--(a);
\draw[ultra thick](e)--(f)--(g)--(h)--(e);
\draw[ultra thick](a)--(e);
\draw[ultra thick](b)--(f);
\draw[ultra thick](c)--(g);
\draw[ultra thick](d)--(h);
\node[left] at (a) {$d$};
\node[right] at (2.6,0,0) {$a$};
\node[right] at (c) {$c$};
\node[left] at (-.1,0,-3) {$d$};
\node[left] at (e) {$d'$};
\node[right] at (2.6,2.5,0) {$a'$};
\node[right] at (g) {$b'$};
\node[left] at (h) {$c'$};
\end{scope}
	\end{tikzpicture}
\caption{${\cal M}^{(abcd)}_{[\{\vec J\}\{\vec K\}]}$ \&   $\Phi(\vec K)$.}
\label{box1}
\end{center}
	\end{subfigure}
\begin{subfigure}[b]{.2\textwidth}
\begin{center}
\begin{tikzpicture}
\draw[line width=2.mm,->,gray] (-1,.0)--(1,.0);
\end{tikzpicture}
\end{center}
\vspace{4.5cm}
\end{subfigure}		
\begin{subfigure}[b]{.4\textwidth}
\begin{center}
\begin{tikzpicture}[scale=1.1]
\coordinate (a) at (0,0,0);
\coordinate (b) at (2.5,0,0);
\coordinate (c) at (2.5,0,-3);
\coordinate (d) at (0,0,-3);
\coordinate (e) at (0,2.5,0);
\coordinate (f) at (2.5,2.5,0);
\coordinate (g) at (2.5,2.5,-3);
\coordinate (h) at (0,2.5,-3);
\shade[fill={gray},top color=gray!60,bottom color=gray!02] (a)--(b)--(c)--(d);
\draw[ultra thick](e)--(f)--(g)--(h)--(e);
\draw[ultra thick](a)--(e);
\draw[ultra thick](b)--(f);
\draw[ultra thick](c)--(g);
\draw[ultra thick](d)--(h);
\node[left] at (a) {$d'$};
\node[right] at (2.6,0,0) {$a'$};
\node[right] at (c) {$b'$};
\node[left] at (-.1,0,-3) {$c'$};
\node[left] at (e) {$h$};
\node[right] at (2.6,2.5,0) {$e$};
\node[above] at (g) {$f$};
\node[above] at (h) {$g$};
\begin{scope}[shift={(0,-2.5)}]
\coordinate (a) at (0,0,0);
\coordinate (b) at (2.5,0,0);
\coordinate (c) at (2.5,0,-3);
\coordinate (d) at (0,0,-3);
\coordinate (e) at (0,2.5,0);
\coordinate (f) at (2.5,2.5,0);
\coordinate (g) at (2.5,2.5,-3);
\coordinate (h) at (0,2.5,-3);
\draw[ultra thick](a)--(b)--(c)--(d)--(a);
\draw[ultra thick,dotted](e)--(f)--(g)--(h)--(e);
\draw[ultra thick](a)--(e);
\draw[ultra thick](b)--(f);
\draw[ultra thick](c)--(g);
\draw[ultra thick](d)--(h);
\node[left] at (a) {$d$};
\node[right] at (2.6,0,0) {$a$};
\node[right] at (c) {$b$};
\node[left] at (-.1,0,-3) {$c$};
\end{scope}
	\end{tikzpicture}
		\vspace{1.3cm}
\caption{$\Phi(\vec J)$}\label{box2}
\end{center}
	\end{subfigure}
\caption{The graphical representation of the Wilson loop eigenvalue equation (\ref{meve}):~ ${\cal M}^{(abcd)}_{[\{\vec J\}\{\vec K\}]}~\Phi(\vec K)= \Phi(\vec J)$ leading to the  Wigner coefficient identities (\ref{boxabn1}): 12j $\times$ 12j =12j.}
\end{figure}
$${M}^{(abcd)}(\vec J, \vec K) = \left(\prod_{l=5,\,6,\,\dots,\,12}
\delta_{j_l,\,k_l} \right)
\frac{\Pi\left(j_1,\; j_2,\; j_3,j_4,\; k_1,\; k_2,\; k_3,\; k_4\right)}{\Pi^4(\frac{1}{2})}$$
Now from (\ref{identity1}) we get 
\begin{align} 
\sum_{\{k\}'} \Pi^2(k_1,k_2,k_3,k_{4}) ~
\left[ \begin{array}{ccccccccc}
\!{j}_{1} \!\!&&\!\! {j}_{2}\!\! &&\!\! j_3\!\!&&\!\!j_4\!\! &\!\!\\
 &\!\!j_5\!\!&& \!\!j_{6} \!\!&& \!\!j_7\!\!&&\!\!j_8\!\!\\
\!{k}_{1}\!\! &&\!\! k_{2}\!\! &&\!\! k_{3}\!\!&&\!\!k_{4}\!\!&\\
\end{array} \right]& 
\left[ \begin{array}{ccccccccc}
\!{k}_{1} \!\!&&\!\! {k}_{2}\!\! &&\!\! k_3\!\!&&\!\!k_4\!\! &\!\!\\
 &\!\!j_5\!\!&& \!\!j_{6} \!\!&& \!\!j_7\!\!&&\!\!j_8\!\!\\
\!{j}_{9}\!\! &&\!\! j_{10}\!\! &&\!\! j_{11}\!\!&&\!\!j_{12}\!\!&\\
\end{array} \right]\nonumber\\
&~~~= \Pi^4 ({\frac{1}{2}})\left[ \begin{array}{ccccccccc}
\!{j}_{1} \!\!&&\!\! {j}_{2}\!\! &&\!\! j_3\!\!&&\!\!j_4\!\! &\!\!\\
 &\!\!j_5\!\!&& \!\!j_{6} \!\!&& \!\!j_7\!\!&&\!\!j_8\!\!\\
\!{j}_{9}\!\! &&\!\! j_{10}\!\! &&\!\! j_{11}\!\!&&\!\!j_{12}\!\!&\\
\end{array} \right]
\label{boxabn1}
\end{align}
As before $\{k\}'$ means 4 triangular constraints over summations on the final momenta,  
i.e.  $k_1,k_2,k_3,k_4$ and the corresponding initial momenta $j_1,j_2,j_3,j_4$ differ by $\pm\frac{1}{2}$ respectively. 
We can also  consider a larger  Wilson loop ${\cal W}_{abfgcda}$. 
We now get 
\bea 
{\cal M}^{(abfgcda)}_{[\{\vec J\}~\{\vec K\}]} & = & \bra{\{\vec J\}} \;{\cal W}_{abfgcda}\; \ket{\{\vec K\}} \nonumber \\ 
& = & 
{M}^{(abfgcda)}(\vec J, \vec K) ~\underbrace{\left[ \begin{array}{ccccccccccccc}
\!\!{j}_{1}\!\! && \!\!{j}_{2}\!\! &&\!\! j_{6}\!\!&&\!\!j_{11}\!\! && \!\!j_7\!\!&&\!\!j_4\!\!\\
\!\! &\!\!j_5\!\!&& \!\!j_{3}\!\! &&\!\! j_{10}\!\!&&\!\!j_{12}\!\!&&\!\!j_3\!\!&&\!\!j_8\!\!\\
\!\!{k}_{1}\!\! && \!\!{k}_{2}\!\! &&\!\! k_{6}\!\!&&\!\!k_{11}\!\! && \!\!k_7\!\!&&\!\!k_4\!\!\\
\end{array} \right]}_{18j ~\text{ coefficient~ of }~2^{\text{nd}}~\text{kind}} \nonumber
\label{boxxvbp} 
\eea 
Like in the previous cases 
\bea 
{M}^{(abfgcda)}(\vec J, \vec K) = 
\delta_{j_3,\,k_3} \delta_{j_5,\,k_5} \delta_{j_8,\,k_8}\delta_{j_9,\,k_9} \delta_{j_{10}, k_{10}} \delta_{j_{12},k_{12}}
\frac{\Pi\left(j_1,j_2,j_4,j_6,j_7,j_{11},k_1,k_2,k_4,k_6,k_7,k_{11}\right)}{\Pi^4(\frac{1}{2})}. 
\nonumber 
\eea 
Now from \ref{identity1} we get following identity for $12j$ coefficients.
\begin{align} 
\sum_{\{k\}'} \Pi(k_1,k_2,k_4,k_6,k_7,k_{11})&
\left[ \begin{array}{ccccccccccccc}
\!\!{j}_{1}\!\! && \!\!{j}_{2}\!\! &&\!\! j_{6}\!\!&&\!\!j_{11}\!\! && \!\!j_7\!\!&&\!\!j_4\!\!\\
\!\! &\!\!j_5\!\!&& \!\!j_{3}\!\! &&\!\! j_{10}\!\!&&\!\!j_{12}\!\!&&\!\!j_3\!\!&&\!\!j_8\!\!\\
\!\!{k}_{1}\!\! && \!\!{k}_{2}\!\! &&\!\! k_{6}\!\!&&\!\!k_{11}\!\! && \!\!k_7\!\!&&\!\!k_4\!\!\\
\end{array} \right]\!\left[ \begin{array}{ccccccccc}
\!{k}_{1} \!\!&&\!\! {k}_{2}\!\! &&\!\! j_3\!\!&&\!\!k_4\!\! &\!\!\\
 &\!\!j_5\!\!&& \!\!k_{6} \!\!&& \!\!k_7\!\!&&\!\!j_8\!\!\\
\!{j}_{9}\!\! &&\!\! j_{10}\!\! &&\!\! k_{11}\!\!&&\!\!j_{12}\!\!&\\
\end{array} \right]\nonumber\\&~~~~~~~~~~~~~~~~=\Pi^4 ({\frac{1}{2}})~\left[ \begin{array}{ccccccccc}
\!{j}_{1} \!\!&&\!\! {j}_{2}\!\! &&\!\! j_3\!\!&&\!\!j_4\!\! &\!\!\\
 &\!\!j_5\!\!&& \!\!j_{6} \!\!&& \!\!j_7\!\!&&\!\!j_8\!\!\\
\!{j}_{9}\!\! &&\!\! j_{10}\!\! &&\!\! j_{11}\!\!&&\!\!j_{12}\!\!&\\
\end{array} \right]
\label{boxabn11}
\end{align}
As in the cases before, the identity (\ref{boxabn11}) can also be generalized to arbitrary spin $s$ representation by replacing $\frac{1}{2}$ by $s$ and modifying the triangular constraints 
appropriately. 
\subsection{SU(2) Toric code model \& topological ground states}
\label{tcm}
In the previous sections we considered the magnetic ground states to get various Wigner coefficient identities. We now extend this method  further and use magnetic ground states with topological charges 
leading to more general class of  identities  parametrized by non-trivial `topological phase factors'.
We consider  exactly solvable SU(2) toric code model 
defined on a two  dimensional torus with periodic boundary conditions in both directions\cite{manatul}. 
The  model has 4 fold degenerate ground states with topological charges $({\mathsf p,\mathsf q}); ~{\mathsf p,\mathsf q}=0,1$. 
In this paper we work with a simple 4 plaquette torus ${\cal T}_2$ to illustrate the ideas. 
 We obtain Wigner coefficient identities with non trivial $(\mathsf p,\mathsf q)$ dependent `topological phase factors'  only for  gauge invariant Wilson lines or Polyakov lines encircling the entire torus ${\cal T}_2$  in any of the two directions. We further show that these  $(\mathsf p,\mathsf q)$ dependent phase factors cancel out from the identities for any local gauge invariant Wilson loops. This is as expected because  one needs to traverse the entire torus to detect the topological charge of the state \cite{kitaev}. 

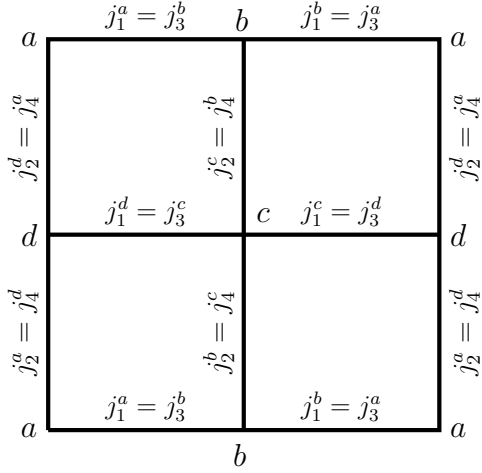
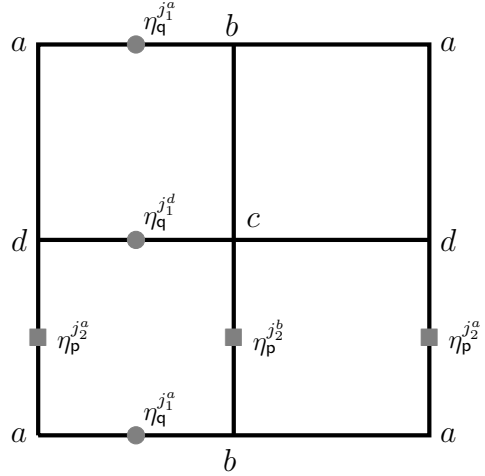
\begin{figure}[t] 
 \begin{subfigure}[]{.5\textwidth}
 	\begin{center}
 		\begin{tikzpicture}[scale=1.3]
 		\draw[ultra thick] (-2,-2) --(2,-2)--(2,2)--(-2,2)--(-2,-2);
 		\draw[ultra thick] (-2,0) --(2,0);
 		\draw[ultra thick] (0,-2) --(0,2);       
 		\node[left]at (-2,-2.){$a$};
 		\node[]at (-0.04,-2.25){$b$};
 		\node[right]at (2,-2.){$a$};
 		\node[right]at (-0.2,2.2){$b$};
 		\node[right]at (2,0.){$d$};
 		\node[left]at (-2,2.){$a$};
 		\node[right]at (2,2.){$a$};
 		\node[left]at (-2,0.){$d$};
 		\node[]at (.2,.2){$c$};
 		\node[above,scale=.85]at (-1,-2){$j_1^a=j_3^b$};
 		\node[above,scale=.85]at (-1,0){$j_1^d=j_3^c$};
 		\node[above,scale=.85]at (-1,2){$j_1^a=j_3^b$};
 		\node[above,scale=.85]at (1,-2){$j_1^b=j_3^a$};
 		\node[above,scale=.85]at (1,0){$j_1^c=j_3^d$};
 		\node[above,scale=.85]at (1,2){$j_1^b=j_3^a$};
\node[above,scale=.9,rotate=90]at (-2,-1){$j_2^a=j_4^d$};
\node[above,scale=.9,rotate=90]at (-2,1){$j_2^d=j_4^a$};
\node[above,scale=.9,rotate=90]at (0,-1){$j_2^b=j_4^c$};
\node[above,scale=.9,rotate=90]at (0,1){$j_2^c=j_4^b$};
\node[below,scale=.9,rotate=90]at (2,-1){$j_2^a=j_4^d$};
\node[below,scale=.9,rotate=90]at (2,1){$j_2^d=j_4^a$};
 		\end{tikzpicture}
 		\caption{Torus ${\cal T}_2$}\label{4plattice}
 	\end{center}
 \end{subfigure}
 \begin{subfigure}[]{.5\textwidth}
 	\begin{center}
 		\begin{tikzpicture}[scale=1.3]
 		\draw[ultra thick] (-2,-2) --(2,-2)--(2,2)--(-2,2)--(-2,-2);
 		\draw[ultra thick] (-2,0) --(2,0);
 		\draw[ultra thick] (0,-2) --(0,2);       
 		\node[left]at (-2,-2.){$a$};
 		\node[]at (-0.04,-2.25){$b$};
 		\node[right]at (2,-2.){$a$};
 		\node[right]at (-0.2,2.2){$b$};
 		\node[right]at (2,0.){$d$};
 		\node[left]at (-2,2.){$a$};
 		\node[right]at (2,2.){$a$};
 		\node[left]at (-2,0.){$d$};
 		\node[]at (.2,.2){$c$};
\draw[fill,gray] (-1,-2) circle(2.5pt);
\draw[fill,gray] (-1,0) circle(2.5pt);
\draw[fill,gray] (-1,2) circle(2.5pt);
\node[diamond,gray,scale=.8](a) at (-2,-1){$\blacksquare$};
\node[diamond,gray,scale=.8](a) at (0,-1){$\blacksquare$};
\node[diamond,gray,scale=.8](a) at (2,-1){$\blacksquare$};
\node[above,scale=.85]at (-.75,-2){$\eta_{\mathsf{q}}^{j_1^a}$};
\node[above,scale=.85]at (-.75,0){$\eta_{\mathsf{q}}^{j_1^d}$};
\node[above,scale=.85]at (-.75,2){$\eta_{\mathsf{q}}^{j_1^a}$};
\node[right,scale=.85]at (-1.9,-1){$\eta_{\mathsf{p}}^{j_2^a}$};
\node[right,scale=.85]at (0.1,-1){$\eta_{\mathsf{p}}^{j_2^b}$};
\node[right,scale=.85]at (2.1,-1){$\eta_{\mathsf{p}}^{j_2^a}$};
 		\end{tikzpicture}
 		\captionof{figure}{topological charges, $\eta_{\mathsf{p}}=(-1)^{2\mathsf{p}},\eta_{\mathsf{q}}=(-1)^{2\mathsf{q}}$}\label{tplattice}
 	\end{center}
 \end{subfigure} 
 		\caption{(a) The 4 plaquette torus ${\cal T}_2$  with plaquettes $\square_{abcd}, \square_{badc}, \square_{cdab},\square_{dcba}$. The 12 Angular momenta are shown on the edges, (b) The topological phase factor $(\eta_{\mathsf p}, \eta_{\mathsf q})$: $\eta_{\mathsf p}$ denoted by {\small \textcolor{gray}{$\blacksquare$}} on the vertical links (ad), (bc), $\eta_{\mathsf q}$ denoted by {\Large \textcolor{gray}{$\bullet$}} on the horizontal links (ab), (dc). Note that these phase factors $\pm 1$ do not change the plaquette magnetic fields.}\label{4platticetp}
 \end{figure}

\begin{figure}
        \begin{subfigure}[b]{.5\textwidth}
        \begin{center}
        \begin{tikzpicture}[scale=.92]
     \draw[ultra thick] (-3,-3) rectangle (3,3);
     \draw[ultra thick] (-1.5,-1.5) rectangle (1.5,1.5);
     \draw[ultra thick] (1.5,-1.5)-- (3,-3);
      \draw[ultra thick] (1.5,1.5)-- (3,3);
        \draw[ultra thick] (-1.5,1.5)-- (-3,3);
          \draw[ultra thick] (-1.5,-1.5)-- (-3,-3); 
\node[above, right]at (-1.5,-1.2){$a$};
\node[above, left]at (1.5,-1.2){$b$};
\node[below, right]at (-1.5,1.15){$d$};
\node[ below]at (1.2,1.45){$c$};
\node[below]at (-3,-3){$a$};
\node[below]at (3,-3){$b$};
\node[above] at (-3,3){$d$};
\node[above]at (3,3){$c$};
\node[above,scale=1.]at (0,3.1){$j^d_1$};
\draw[fill,gray] (0,3) circle(4pt);
\node[above,scale=1.]at (0,1.5){$j^c_1$};
\node[below,scale=1.]at (0,-1.6){$j^a_1$};
\draw[fill,gray] (0,-1.5) circle(4pt);
\node[below,scale=1.]at (0,-3){$j^b_1$};
\node[above,scale=1.,rotate=90]at (-3,0){$j^d_2$};
\node[above,scale=1.,rotate=90]at (-1.6,0){$j^a_2$};
\node[gray,scale=1.]at (-1.5,0){$\blacksquare$};
\node[below,scale=1.,rotate=90]at (1.5,0){$j^c_2$};
\node[below,scale=1.,rotate=90]at (3.1,0){$j^b_2$};
\node[gray,scale=1.]at (3,0){$\blacksquare$};
\node[scale=1.]at (-2.3,-2.){$j^a$};
\node[scale=1.]at (2.5,-1.8){$j^b$};
\node[scale=1.]at (-2.3,1.8){$j^d$};
\node[scale=1.]at (2.3,1.8){$j^c$};
        \end{tikzpicture}
\caption{The 12j amplitude $\Phi_{\mathsf{p},\,\mathsf{q}}(\vec J)$}\label{tc12j}
\end{center}
        \end{subfigure}
        \hspace{-.5cm}
                \begin{subfigure}[b]{.5\textwidth}
        \begin{center}
        \begin{tikzpicture}[scale=.8]
       \draw[ultra thick] (xy polar cs:angle=0,radius=2.2)--(xy polar cs:angle=60,radius=2.2);
        \draw[ultra thick] (xy polar cs:angle=60,radius=2.2)--(xy polar cs:angle=120,radius=2.2);
           \draw[ultra thick] (xy polar cs:angle=120,radius=2.2)--(xy polar cs:angle=180,radius=2.2);
                  \draw[ultra thick] (xy polar cs:angle=180,radius=2.2)--(xy polar cs:angle=240,radius=2.2);
                    \draw[ultra thick] (xy polar cs:angle=240,radius=2.2)--(xy polar cs:angle=300,radius=2.2);
                           \draw[ultra thick] (xy polar cs:angle=300,radius=2.2)--(xy polar cs:angle=0,radius=2.2); 
  \draw[ultra thick] (xy polar cs:angle=0,radius=4)--(xy polar cs:angle=60,radius=4);
        \draw[ultra thick] (xy polar cs:angle=60,radius=4)--(xy polar cs:angle=120,radius=4);
           \draw[ultra thick] (xy polar cs:angle=120,radius=4)--(xy polar cs:angle=180,radius=4);
                  \draw[ultra thick] (xy polar cs:angle=180,radius=4)--(xy polar cs:angle=240,radius=4);
                    \draw[ultra thick] (xy polar cs:angle=240,radius=4)--(xy polar cs:angle=300,radius=4);
                           \draw[ultra thick] (xy polar cs:angle=300,radius=4)--(xy polar cs:angle=0,radius=4); 
 \draw[ultra thick] (xy polar cs:angle=0,radius=2.2)--(xy polar cs:angle=0,radius=4);
        \draw[ultra thick] (xy polar cs:angle=60,radius=2.2)--(xy polar cs:angle=60,radius=4);
           \draw[ultra thick] (xy polar cs:angle=120,radius=2.2)--(xy polar cs:angle=120,radius=4);
                  \draw[ultra thick] (xy polar cs:angle=180,radius=2.2)--(xy polar cs:angle=180,radius=4);
                    \draw[ultra thick] (xy polar cs:angle=240,radius=2.2)--(xy polar cs:angle=240,radius=4);
                           \draw[ultra thick] (xy polar cs:angle=300,radius=2.2)--(xy polar cs:angle=300,radius=4);
\node[scale=1.] at (xy polar cs:angle=90,radius=1.4) {$k_1^a$};                                
\node[scale=1.] at (xy polar cs:angle=150,radius=1.4) {$k_2^b$};                                
\node[scale=1.] at (xy polar cs:angle=210,radius=1.4) {$k^d$};                                
\node[scale=1.] at (xy polar cs:angle=270,radius=1.4) {$k_1^d$};                                
\node[scale=1.] at (xy polar cs:angle=330,radius=1.4) {$k_2^b$};                                
\node[scale=1.] at (xy polar cs:angle=30,radius=1.4) {$k^b$};                                
\node[scale=1.] at (xy polar cs:angle=90,radius=4) {$j_1^a$};                                
\node[scale=1.] at (xy polar cs:angle=150,radius=4) {$j_2^a$};                                
\node[scale=1.] at (xy polar cs:angle=210,radius=4) {$j^d$};                                
\node[scale=1.] at (xy polar cs:angle=270,radius=4) {$j_1^d$};                                
\node[scale=1.] at (xy polar cs:angle=330,radius=4) {$j_2^b$};                                
\node[scale=1.] at (xy polar cs:angle=30,radius=4) {$j^b$};                                
\node[scale=1.] at (xy polar cs:angle=10,radius=3) {$j_1^b$};                                
\node[scale=1.] at (xy polar cs:angle=67,radius=3) {$j_2^c$};                                
\node[scale=1.] at (xy polar cs:angle=128,radius=3) {$j^a$};                                
\node[scale=1.] at (xy polar cs:angle=172,radius=3) {$j_1^c$};                                
\node[scale=1.] at (xy polar cs:angle=232,radius=3) {$j_2^d$};                                
\node[scale=1.] at (xy polar cs:angle=290,radius=3) {$j^c$};
        \end{tikzpicture}
\caption{The 18j  matrix elements ${\cal M}^{abcd}_{[\{\vec J \} \, \{\vec K\}]}$} \label{18j}
\end{center}
        \end{subfigure}
        \hspace{-.2cm}
        \caption{ The  ampltiude $\Phi_{({\mathsf p},\,{\mathsf q})} (\vec J)$ and the matrix elements $ {\cal M}^{abcd}_{[\{\vec J \} \, \{\vec K\}]}$ on the torus ${\cal T}_2$ (a) The diagram implies: $ \Phi_{({\mathsf p},\,{\mathsf q})} (\vec J) = (-1)^{2\mathsf{p}(j^a_2+j^b_2)} (-1)^{2\mathsf{q}(j^d_1+j^d_1)}\Phi_{(0,\,0)}(\vec J)$,~ (b) The matrix elements $ {\cal M}^{abcd}_{[\{\vec J \} \, \{\vec K\}]}$do not depend on the topological phases.} 
\end{figure}
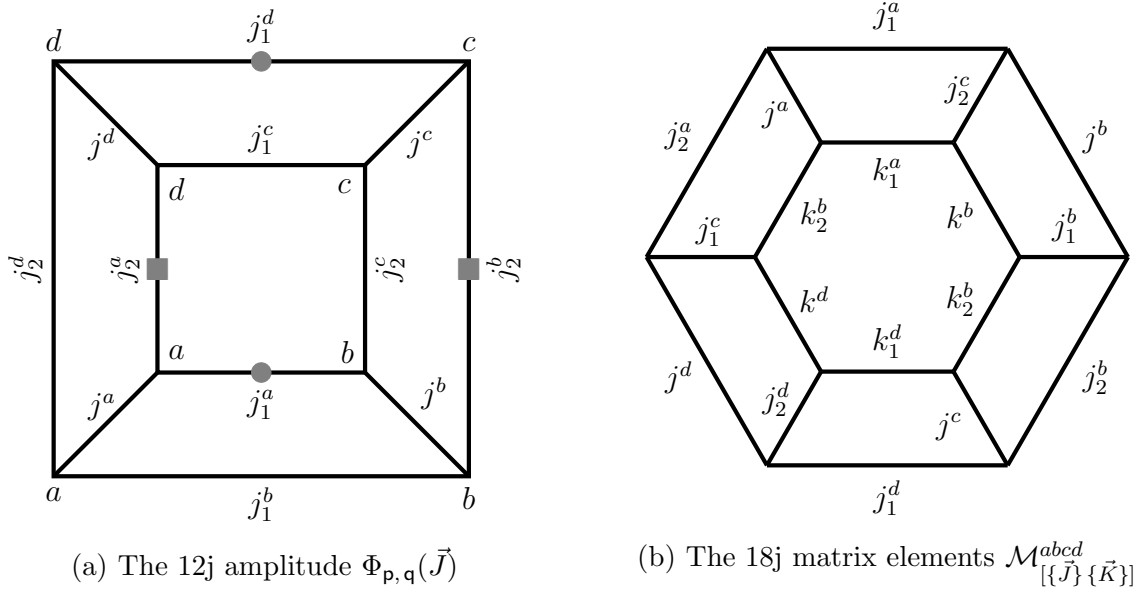

We consider a small 4 plaquette lattice on torus as shown in Figure \ref{4plattice}.  The 4 plaquettes are denoted by ${\cal U}_p \equiv \square_{abcd}, \square_{badc}, \square_{cdba}, \square_{dcba}$. 
The SU(2) Kitaev toric code Hamiltonian is
\begin{equation}
H = A \sum_{n} A_n + B \sum_p B_p. 
\label{suntchx} 
\end{equation}
where $A$ and $B$ are positive constants,   $n$ and $p$ denote the sites and plaquettes and 
\bea  
A_n \equiv  \sum_{a=1}^{3} \;{\cal G}^a(n) \;{\cal G}^a(n); ~~~~~~~~~~
{\cal B}_p \equiv \Big(1- \frac{1}{2}\; \text{Tr}~\mathcal{U}_p. 
\Big). 
\label{abes1} 
\eea
 The Gauss law generators ${\cal G}^a(n)$ and the plaquette terms ${\cal U}_p$ have been defined in  (\ref{gtg}) and  (\ref{sunph}) respectively. 
To construct the spin network on a torus, it is 
convenient to label the   4 angular momenta 
around a  vertex $v (=a,b,c,d)$ in a counter clockwise direction as $j^v_1,j^v_2,j^v_3, j^v_4$ as shown in Figure \ref{vertex}. As each edge is shared by two vertices there is double counting and we identify: 
\begin{align}
\begin{aligned}
&j_1^a \equiv j_3^b, ~~~~~~j_2^a \equiv j_4^d, ~~~~~~j_1^b \equiv j_3^a, ~~~~~~ j_2^d \equiv j_4^a, \\
&j_2^b \equiv  j_4^c,~~~~~\,~ j_2^c\equiv j_4^b ,~\,~~~~~ j_1^c \equiv j_3^d,~~~~~~ j_1^d 
\equiv j_3^c. 
\label{ident} 
\end{aligned}
\end{align}
Thus there are 8 angular momenta on 
 8  edges or links on the 4 plaquette torus ${\cal T}$.  
We combine two set of angular momentum $j^v_1, j^v_2$ to get $j^v_{12}$ and $j^v_3, j^v_4$ to get $j^v_{34}$ and then $j^a_{12}, j^v_{34}$ are add to get the gauge invariant  
$0$ angular momentum states. We set $j^v_{12}=j^v_{34}= j^v$, now the  spin network states\footnote{Spin network states  at site $v=1,2,...,L^2$ is given by
	\bea 
	\ket{\vec J_v} \equiv \ket{ j^v_1, j^v_2,j^v_3,j^v_4,j^v}  = \!\!\!
	\sum_{\text{all}\; m}\! A^{j^v}_{m^v} 
 C_{j_1^v,m^v_1, j_2^v,m^v_2}^{j^v, m^v}  
	C_{j_3^v,m^v_3 j_4^v m^v_4}^{j^v, -m^v}  
	\ket{j_1^v,m_1^v} \otimes 
	\ket{j_2^v,m_2^v} \otimes \ket{j_3^v,m_3^v}\otimes\ket{j_4^v,m_4^v}\equiv\ket{j_1^v,j_2^v,j^v}.\nonumber  
	\eea Here $A^j_m \equiv \frac{(-1)^{(j-m)}}{\sqrt{(2j+1)}}$.} 
at any vertex are characterized by a set of five angular momenta $\ket{\vec J_v} = \ket{j^v_1, j^v_2,j^v_3,j^v_4,j^v}$ with the identifications in (\ref{ident}). 
Infact, the 8 identifications (\ref{ident}) show that all   angular momenta in the 3 and 4 directions (r.h.s of 
(\ref{ident})) can be replaced by the neighbouring site angular momenta in 1 and 2 directions. 
Therefore, the spin network states at a vertex $v$
can be  represented by $\ket{\vec J_v} \equiv \ket{j_1^v,j_2^v,j^v}$. Thus the gauge invariant  states on ${\cal T}_2$ are characterized by 12 angular momenta (8 on the 8 edges and 4 on the 4 vertices). 
We expand the ground state $\ket{\psi_0}$,
satisfying 
${\cal G}^a\;\ket{\psi_0}=0$, in the  spin network basis 
\bea 
\ket{\psi_0} = \sum_{\{\vec J_a,\,\vec J_b,\,\vec J_c,\,\vec J_d\}}
\underbrace{\Phi \left(\vec J_a,\vec J_b,\vec J_c,\vec J_d\right)}_{\text{amplitude}}  ~
\underbrace{\ket{\vec J_a} \otimes \ket{\vec J_b} \otimes \ket{\vec J_c} \otimes \ket{\vec J_d}}_{\text{spin network states on ${\cal T}_2$}}. 
\label{fwf} 
\eea 
In  Appendix A,  we compute  the amplitudes 
$\Phi$ by choosing the standard pure gauge conditions 
on all the 8 edges to get ${\cal B}_p=0$  and then integrating over the SU(2) gauge parameters 
on the 4 vertices to make $\ket{\psi_0}$  gauge invariant. As expected from the diagram in Figure \ref{tc12j}, the amplitudes  are 12j Wigner coefficients of 2nd kind:  
\bea 
\Phi\left(\vec J_a,\vec J_b,\vec J_c,\vec J_d\right)= \;\Pi (\{ \vec{J}\})
\begin{bmatrix}\; j_2^a=j_4^d\!\! && \!\! j_1^a=j_3^b\!\! && \!\! j_2^c=j_4^b\!\! && \!\! j_{1}^c=j_3^d\!\! &&\\
&\!\! j^a \!\! && \!\! j^b \!\! && \!\! j^c\!\! && \!\! j^d\!\! \!\! \!\! \\
\;j_2^d =j_4^a\!\! &&\!\! j_1^b =j_3^a\!\! && \!\! j_2^b =j_4^c\!\! && \!\! j_1^d=j_3^c\!\! &&\\
\end{bmatrix}\label{bmatrix1}
\eea
We can now construct the topological ground states (see Appendix A) by modifying the pure gauge configurations (\ref{pgc}) with the following $Z_2$ factors\footnote{Equivalently, we can  modify the pure gauge configurations  
on the links (da), (cb) by $(-1)^{\mathsf p}$ and (ba), (cd) by $(-1)^{\mathsf q}$. They all leave plaquette magnetic fields unchanged.} $(-1)^{\mathsf p}, ~(\mathsf p = 0,1)$ 
on the links $(ad)$ and $(bc)$, (b) $(-1)^{\mathsf q} ~
(\mathsf q =0,1)$ 
on the links $(ab)$ and $(cd)$. The magnetic field ${\cal B}_p$ on each plaquette remains unchanged but the amplitude $\Phi$ gets replaced: $\Phi\left(\vec J_a,\vec J_b,\vec J_c,\vec J_d\right)\rightarrow \Phi_{({\mathsf p},\,{\mathsf q})} \left(\vec J_a,\vec J_b,\vec J_c,\vec J_d\right)$ where, 
\bea 
\Phi_{({\mathsf p},\,{\mathsf q})}\left(\vec J_a,\vec J_b,\vec J_c,\vec J_d\right)
= (-1)^{2 {\mathsf p} \left(j_2^a+j_2^b\right)} ~(-1)^{2 {\mathsf q}\left(j_1^a+j_1^d\right)} ~~\Phi\left(\vec J_a,\vec J_b,\vec J_c,\vec J_d\right), ~~~\mathsf p, \mathsf q =0,1.  
\label{tp}
\eea 
Now we can  generalized identities (\ref{identity1}) for topological charge and write down following identities;
\bea \label{identity1tp}
\sum_{\{\vec k\}'}\; {\cal M}^{~({\cal C})}_{[\{\vec J\}\,\{\vec K\}]}~ \Phi_{({\mathsf p},\,{\mathsf q})} (\vec K\;) 
= \Phi_{({\mathsf p},\,{\mathsf q})}  (\vec J\;)
\label{mevetp} 
\eea 
For ${\mathsf p}={\mathsf q}=0$ these identities coincide 
with the standard identities  (\ref{identity1}) for $\omega=0$. As before, $\{\vec k\}'$ means that the summations  over $\{\vec k\}$ are restricted by the representation of the 
loop operator involved in computing ${\cal M}$. 
We now consider the Wilson loop to be the plaquette operator 
$\square_{abcd}$ in $s=\frac{1}{2}$ representation:
\bea 
{\cal M}^{abcd}_{[\{\vec J\}\,\{\vec K\}\,]}=  \frac{1}{2}\; 
\langle {\vec K_d}| \otimes \langle {\vec K_c}| \otimes \langle {\vec K_b}|\otimes \langle {\vec K_a}| \;{\text Tr}~\square_{abcd} \;\ket{\vec J_a} \otimes \ket{\vec J_b} \otimes \ket{\vec J_c} \otimes \ket{\vec J_d}
\eea 
are  $18j$ Wigner coefficients of second kind and can be represented by  
a ribbon diagram shown in Figure-\ref{18j}:
\begin{align}
{\cal M}^{~abcd}_{[\{\vec{J}\},\,\{\vec{K}\}]}=M^{abcd}(\vec J, \vec K)~ \left[ \begin{array}{cccccccccccc}
\!\! j^a_1\!\! &&\!\! j_2^a\!\! &&\!\!  j^d\!\!  && \!\! j_1^d\!\! && \!\! j_2^b\!\!  &&\!\!  j^b\!\! &\!\! \\
\!\! &\!\! j^a\!\! & &\!\! j_1^c\!\! &&\!\! j_2^d\!\! && \!\! j^c&&\!\! j^b_1\!\!  &&\!\!  j_2^c\!\! \\
\!\! k^a_1\!\! &&\!\! k_2^a\!\! && \!\! k^d \!\! && \!\! k_1^d\!\! && \!\! k_2^b \!\! && \!\! k^b\!\! &\\
\end{array}\right]~\{\vec j, \vec k, \frac{1}{2}\}.
\label{18jme}
\end{align} 
\bea 
M^{abcd}(\vec J, \vec K) 
= \delta_{j^a,\,k^a} \; \delta_{j^c_1,\,k^c_1} \;
\delta_{j_2^d,\,k_2^d} \; \delta_{j^c,\,k^c} \; \delta_{j_1^b,k_1^b}\, \delta_{j_2^c,\, k_2^c}~
\frac{\Pi\left(j_1^a,j_2^a,j^d, j_1^d, j_2^b, j^b,k_1^a,k_2^a,k^d, k_1^d, k_2^b, k^b\right)}{\Pi^4(\frac{1}{2})}. 
\nonumber 
\eea 
The constraints in $\{\vec j,\vec k,\frac{1}{2}\}$ imply 
that the initial and the final angular momenta in (\ref{18jme}) differ by $\pm\frac{1}{2}$. 
Similarly, we can write these matrix elements for other plaquette operators. The action of ${\text Tr}\; \square_{abcd}$ in an arbitrary spin $s$ representation can 
be obtained by the replacements: 
$\{\vec j,\vec k,\frac{1}{2}\} \rightarrow \{\vec j,\vec k, s \}$ and $\Pi(\frac{1}{2}) \rightarrow  \Pi(s) $.
\subsubsection{Wigner Identities for 12j-18j and 18j-18j Coefficients}

Using above matrix elements obtained above we get;
\begin{align}
\begin{aligned}
\sum_{\{k\}'}\, \Pi^2(\vec{k})\left[ \begin{array}{cccccccccccc}
\!\! j^a_1\!\! &&\!\! j_2^a\!\! && \!\! j^d\!\!  &&\!\!  j_1^d\!\! && \!\! j_2^b \!\! && \!\! j^b&\!\! \\
&\!\! j^a\!\! & &\!\! j_1^c\!\! &&\!\! j_2^d\!\! &&\!\!  j^c\!\! &&\!\! j^b_1\!\!  &&\!\!  j_2^c\!\! \\
\!\! k^a_1\!\! &&\!\! k_2^a\!\! &&\!\!  k^d\!\!  && \!\! k_1^d\!\! && \!\! k_2^b\!\!  &&\!\!  k^b&\\
\end{array}\right]&\left[\begin{array}{cccccccccccc}
\!\! k^a_2\!\! &&\!\! k_1^a\!\! &&\!\!  j_2^c\!\!  &&\!\!  j_1^c\!\!  \\
&\!\! j^a\!\! &&\!\! k^b\!\! &&\!\! j^c\!\! && \!\! k^d \\
\!\! j^d_2\!\! &&\!\! j_1^b\!\! &&\!\!  k_2^b\!\!  &&\!\!  k_1^d\!\!  \\ \end{array}\right]\\ 
&~~=\Pi^4 ({\frac{1}{2}})~\left[\begin{array}{cccccccccccc}
\!\! j^a_2\!\! &&\!\! j_1^a\!\! && \!\! j_2^c\!\!  &&\!\!  j_1^c\!\!  \\
&\!\! j^a\!\! &&\!\! j^b\!\! &&\!\! j^c\!\! && \!\! j^d\!\!  \\
\!\! j^d_2\!\! &&\!\! j_1^b\!\! &&\!\!  j_2^b\!\!  && \!\! j_1^d\!\!  \\ \end{array}\right]
\end{aligned} \label{ntpf}
\end{align} 
Note that the topological phase factors cancel out on the 
two sides of (\ref{ntpf}). This can be seen as follows. First note that the matrix elements are computed in the spin network basis and therefore do not depend on the topological charges ($\mathsf{p},\mathsf{q}$). The topological phase appear in (\ref{mevetp}) only through the amplitudes on both sides. The phase factor on the left hand side of (\ref{ntpf}) is $(-1)^{2 {\mathsf p}\; \left(k_2^a+k_2^b\right)} ~(-1)^{2 {\mathsf q}\; \left(k_1^a+k_1^d\right)}$. We first focus on the $\mathsf p$ dependent phase. The 4 triangular constraints present in the matrix element ${\cal M}^{~abcd}_{[\{\vec J\}, \{\vec K\}]}$ in (\ref{18jme}) are  
$\{k_2^a,j_2^a,\frac{1}{2}\}, \{k_2^b,j_2^b,\frac{1}{2}\}$ 
and $ \{k_1^a,j_1^a,\frac{1}{2}\}, \{k_1^d,j_1^d,\frac{1}{2}\}$. This implies  $$(-1)^{k_2^a+k_2^b} = (-1)^{j_2^a+j_2^b}$$ and  $(-1)^{k_1^a+k_1^d} = (-1)^{j_1^a+j_1^d}$. Infact, this cancellation remains valid even if we had chosen the plaquette operator in an arbitrary spin $s$
 representation. We now consider Wigner coefficients identities which get modified by the topological phases.  
\subsubsection{ Non-contractible Wilson loops and Topological sectors}
In order to obtain non-trivial identities from the topological ground states, we need to  consider  non-contractible  operators and their eigenvalue equations.
These operators, known as Wilson lines or Polyakov lines, are defined as path ordered product of the flux operator along a non-contractible curve \cite{manatul, kitaev}. 
On the small torus ${\cal T}_2$ with 4 plaquettes,  we have two horizontal Wilson lines $W _{aba}, W_{dcd}$ and two vertical Wilson lines $W_{ada}, W_{bcb}$;
\begin{align}
W_{ada}=&\, U_{ad} \,U_{da},~~~~~~~~~~~~~~~ W_{bcb}=\, U_{bc} \,U_{ca}\\
W_{aba}=&\, U_{ab} \,U_{ba},~~~~~~~~~~~~~~~ W_{dcd}=\, U_{dc} \,U_{cd}
\end{align}
Topologically non-trivial ground state are given by (see Appendix A )
\bea 
 \ket{\psi_0}_{(\mathsf p,\mathsf q)} = \sum_{\{\vec J\}} ~
\Phi_{(\mathsf p, \mathsf q)} \left(\vec J_a,\vec J_b,\vec J_c,\vec J_d\right)~~ \prod_{v=a,b,c,d} \otimes  ~\ket{\;j_1^v,j_2^v,j^v}.
\label{gsne} 
\eea 
Here $\Phi_{(\mathsf p, \mathsf q)}$ are given in  (\ref{bmatrix1}) and (\ref{tp}). The  matrix elements  of Wilson line $W_{ada}$ in  $s=\frac{1}{2}$ representation are
\bea 
{\cal M}^{(ada)}_{[\{\vec J\}~\{\vec K\}\,]} & = &\frac{1}{2} \bra{\vec J\,} \;W^{ada}\; \ket{\vec K\,} \nonumber \\ 
& = &\, 
{M}^{(ada)}(\vec J, \vec K) 
\left[ \begin{array}{ccccccccc}
\!\!{j}^a\!\! && \!\!{j}_{2}^a\!\! && \!\!j^d\!\!&&\!\!j_2^d\!\! &\\
 &\!\!j_1^a\!\!&& \!\!j_{1}^d\!\! &&\!\! j_1^d\!\!&&\!\!j_1^b\!\\
\!\!{k}^a\!\! && \!\!{k}_{2}^a \!\!&& \!\!k^d\!\!&&\!\!k_2^d \!\!&\\
\end{array} \right]
\{\vec j_l,\,\vec k_l,\,\frac{1}{2} \} \nonumber
\label{wilson1} 
\eea 
where \bea 
M^{ada}(\vec J, \vec K) 
= \delta_{j^a_1,\,k^a_1} \;\delta_{j^b_1,\,k^b_1}\;\delta_{j^b_2,\,k^b_2}\;\delta_{j^b,\,k^b}\;\delta_{j^c_1,\,k^c_1}\;\delta_{j^c_2,\,k^c_2}\;\delta_{j^c,\,k^c}\;\delta_{j^d_1,\,k^d_1}
\frac{\Pi\left(j^a,\,j_2^a,\,j^d,\, j_2^d,\,k^a,\,k_2^a,\,k^d,\, k_2^d,\right)}{\Pi^4(\frac{1}{2})}. 
\nonumber 
\eea 
Now from (\ref{identity1tp}) we get following identities for $12j$ coefficients
\begin{align} 
\begin{aligned}
\sum_{\{k\}'} (-1)^{2\mathsf{p}\, k_2^a}\,\Pi^2(k^a,k_2^a,k^d,k_{2}^d)& ~
\left[ \begin{array}{ccccccccc}
\!{j}^a \!\!&& \!\!{j}_{2}^a\!\! && \!\!j^d&&\!\!j_2^d\! &\\
 \!&\!\!j_1^a\!\!&& \!\!j_{1}^c\!\! &&\!\! j_1^d\!\!&&\!\!j_1^b\!\\
\!{k}^a\!\! &&\!\! {k}_{2}^a \!\!&& \!\!k^d\!\!&&\!\!k_2^d &\!\\
\end{array} \right]\left[ \begin{array}{ccccccccc}
\!\!{k}_{2}^a \!\!&&\!\! {j}_{1}^a\!\! &&\!\! j_2^c\!\!&&j_1^c\!\! &\\
 \!\!& \!\!k^a \!\!&&  \!\!j^b \!\! &&  \!\! j^c \!\!&& \!\!k^d\!\!\\
\!\!{k}_{2}^d  \!\!&&  \!\!{j}_{1}^b \!\! && \!\! j_2^b \!\!&& \!\!j_1^d \!\! &\\
\end{array} \right]\\&~~~~~~=(-1)^{2\mathsf{p}\, j_2^a}\,\Pi^4 ({\frac{1}{2}})\left[ \begin{array}{ccccccccc}
\!\!{j}_{2}^a \!\!&&\!\! {j}_{1}^a\!\! &&\!\! j_2^c\!\!&&j_1^c\!\! &\\
 \!\!& \!\!j^a \!\!&&  \!\!j^b \!\! &&  \!\! j^c \!\!&& \!\!j^d\!\!\\
\!\!{j}_{2}^d  \!\!&&  \!\!{j}_{1}^b \!\! && \!\! j_2^b \!\!&& \!\!j_1^d \!\! &\\
\end{array} \right]
\label{boxada1}
\end{aligned}
\end{align}
Note that the phase factors present on the links $(ab), (dc)$ and $(bc)$ cancel out as the corresponding $js$ do not change. The only contribution comes from the topological phase present on the link $(ad)$ carrying $j_2^a$ flux. 
Further, the trivial  ${\mathsf{p}}=0$  identities in 
(\ref{boxada1}) are exactly same \footnote{The equation (\ref{boxada1}) mathces exactly  with equation (\ref{boxabn1}) after the following replacements  $j^a\rightarrow j_1,j_1^a\rightarrow j_5,j_2^a\rightarrow j_2,j^b\rightarrow j_9,j_1^b\rightarrow j_8,j_2^b\rightarrow j_{12},j^c\rightarrow j_{11},j_1^c\rightarrow j_6,j_2^c\rightarrow j_{10},j^d\rightarrow j_3,j_1^d\rightarrow j_9,j_2^d\rightarrow j_4; k^a\rightarrow k_1,k_2^a\rightarrow k_2,k^d\rightarrow k_3,k_2^d\rightarrow k_4$ and using the symmetry properties of 12j coefficients.}  as the ones obtained 
form the single cube lattice in  section-\ref{box}. It is interesting that  the same identities are obtained by analysing gauge theory on two very different underlying lattices with different  structure of spin networks, the first on a cube and the second on a torus.

\section{Summary \& Discussion}

We considered SU(2) lattice gauge theory in 2 and 3 space dimensions on finite lattices and the associated physical spin network Hilbert spaces. We obtained various 3nj Wigner coefficient identities by analysing  Wilson loop 
and   Wilson line (toric code) operators and their eigenvalue equations in the SU(2) spin network Hilbert space. All identities 
are of the form: $$\sum_{\{\vec K\}'}\;\underbrace{~{\cal M}_{[\{\vec J\}\,\{\vec K\}]}~}_{\text{3nj}} \;\underbrace{~\Phi_\omega(\vec J)~}_{3mj} = \cos  \omega \; \underbrace{~\Phi_\omega(\vec J)~}_{3mj},$$   

\begin{itemize} 
	\item In this work we only considered $\omega=0$ cases on finite small lattices.  The value of $n$ is the length of the Wilson loops 
	or Wilson lines, the value of $m$ depends on the dimensions and size of the lattice.  
	\item The number of summations involved is the number of links 
in the Wilson loop or line (toric code).  The range in the summation can be increased or decreased by considering Wilson loops or lines in the higher or lower spin representations respectively.  Generally, identities involving more than 3 summations are not found in the literature \cite{varsha,yutsis}. In this work we have considered cases upto 6 summations. We can also iterate the above eigenvalue equations multiple times (see equation (\ref{identity1aa})) to obtain more general results. Similarly, $\omega \neq 0$ and higher SU(N) cases will be interesting to analyze.  
	\item The SU(2) toric code Hamiltonian (\ref{suntchx}) is exactly solvable as all the terms present in it mutually commute. Therefore, its eigenvalue equations can also be used 
	to get general identities with additional topological phase factors. 
	\end{itemize} 
\section{Appendix}
\appendix
\section{Spin Network Amplitude on Tetrahedron}
In this section, we will calculate the amplitudes for the  magnetic ground state expanded in the  spin network Hilbert space. 
The final results are used in all the cases discussed 
in this work. We will work out the details for the tetrahedron (\ref{tetraamplitude}) and the torus  where the amplitude gets modified by topological phases (\ref{tp}). These amplitude can be fixed using properties of ground states namely ${\cal B}_p\; \ket{\psi_0}=0$ and ${\cal G}^a_n\;\ket{\psi_0}=0$. We begin with states that are magnetically ordered, ${\cal B}_p\; \ket{\psi_0}=0, ~\forall p$ and then ensure ${\cal G}^a_n \; \ket{\psi_0}=0, ~\forall n$ by demanding invariance under SU(2) gauge transformations (\ref{gts}). 
 First, we observe that eigenstates of magnetic fields are necessarily eigenstates of individual links or flux operators $U_{\alpha\beta}$. 
As  eigen-values of the flux operators are SU(2) matrices, we define SU(2) group manifold 
$S^3$ on every link $l=1,2,\dots , 6$ : 
\bea 
Z(l) = \begin{bmatrix} ~z_1(l) && z_2(l)\\
-z_2^*(l) && z_1^*(l) \\
\end{bmatrix}  ~~~~|z_1(l)|^2 + |z_2(l)|^2 =1;~~ (z_1,z_2) \in S^3. 
\eea 
The these holonomy operators  commute with each other  
$[U_{\alpha\beta}(l), U_{\gamma\delta}(l')] =0,~ \forall l,l' $. Therefore 
we can diagonalize all of them simultaneously. We define the magnetic eigenstates to be eigenstates of 
each link operator: 
\bea 
U_{\alpha\beta}(l) ~|Z(l) \rangle = Z_{\alpha\beta}(l) ~|Z(l)\rangle.
\eea 
The eigenvectors can be expanded in terms of angular momentum states with Wigner D functions as Fourier coefficients.
\bea 
\ket{Z(l)} \equiv |z_1(l),z_2(l)\rangle  = \frac{1}{4\pi} \sum_{j=0}^\infty \sqrt{(2j+1)}\sum_{m_\pm} D^{~j}_{m_+~m_-}(Z (l)) 
\ket{j~m_+}_l\otimes \ket{j~m_-}_l. 
\eea 
To obtain completely ordered  states, $B_p=0$ or equivalently ${\cal W}(p)=1$, we choose pure gauge conditions
on every link and write 
\bea 
Z(l)\equiv Z(n,\hat i)  = \xi(s)~ \xi^\dagger(s').
\label{pgc} 
\eea   
Here $\xi_s$ are the SU(2) matrices defined on the site $s$ . The magnetic eigenvalue equations now decouple into left and right parts: 
\bea 
\ket{Z(l)= Z(l)} = \frac{1}{4\pi} \sqrt{(2j+1)}\sum_{j=0}^{\infty} ~\sum_{\bar m=-j}^{j} ~\ket{\xi(n)}^j_{\bar m}  \otimes \ket{\xi(s')}^{j}_{\bar m}.  
\label{lrp} 
\eea 
The states at the left and the right vertices of the link $l$ in (\ref{lrp}) are  called vertex states and defined as \cite{manatul}; 
\bea 
 \underbrace{\ket{\xi(s)}^j_{\bar m} =\sum_{m_+} D^{~j}_{m_+~\bar m}(\xi(s)) ~\ket{j,m_+}}_{\text{state at left vertex}}, ~~ ~~
\underbrace{\ket{\xi(s')}^j_{\bar m}= \sum_{m_-}  D^{~j}_{\bar m~ m_-} (\xi(s')) ~\ket{j,m_-}}_{\text{state at right vertex}}.
\label{wdms} 
\eea 
 Using pure gauge conditions the completely ordered states on tetrahedron can be written as;
\begin{align}
\begin{aligned}
\ket{\psi}=&\sum_{\text{all}\, j} \sum_{\text{all} \,m\,,n} \prod_{s\in {\cal T}}\;\; \left(\ket{\xi(s)}^{j_1}_{m_1}\,\otimes\ket{\xi(s)}^{j_2}_{m_2}\,\otimes\ket{\xi(s)}^{j_3}_{m_3} \right)
\end{aligned}
\end{align}
Now we can integrate over $\xi(s), s=a,b,c,d$ to get the ground states which satisfies conditions (\ref{ab=0}).
\bea \label{gtd} 
\ket{\psi_0} \!\! &=&  \sum_{\text{all }j}
\sum_{\text{all} ~m}\prod_{s \in {\cal T}} \left\{\int_{S^3} d^2\mu(\xi(s))~  
\ket{\xi(s)}^{j_1}_{m_1}\otimes \ket{\xi(s)}^{j_2}_{m_2}\otimes\ket{\xi^\dagger(s)}^{j_3}_{m_3} \right\}, \nonumber \\
&=& \sum_{\{\vec J\}} ~\Phi\left [\vec J\;\right] ~~ \prod_{s} \otimes  \underbrace{\ket{j_1,j_2,j_3}_s}_{\text{loop state at site} s}\nonumber \\
&=&  \sum_{\text{all }j} \underbrace{ {\Pi(j_1,j_2,j_3,j_4,j_5,j_6)}\left\{ \begin{array}{ccc}
\!\!j_1&j_2& j_3\!\!\\
\!\!j_4&j_5& j_6\!\! 
\end{array} \right\}}_{\equiv \Phi(\vec J)}\;\ket{j_1j_2j_3}\otimes \ket{j_3j_4j_5}\otimes\ket{j_2j_4j_6}\otimes\ket{j_1j_5j_6} 
\label{lrv} 
\eea 
In the above calculation we have used 
\begin{enumerate}
\item Properties of D-functions which implies
\begin{align}
\int_{S^3} d \mu \,(\xi(s))\; \ket{\xi(s)}^{j_1}_{m_1} \ket{\xi(s)}^{j_2}_{m_2} \ket{\xi(s)}^{j_3}_{m_3} =\sqrt{\Pi(j_1,j_2,j_3)} \left( \begin{array}{ccc}
\!\!j_1&j_2& j_3\!\!\\
\!\!m_1&m_2& m_3\!\! 
\end{array} \right)\;\ket{j_1j_2j_3}
\end{align} 
\item   Loop states $\ket{j_1,j_2,j_3}$ which are given by 
\begin{equation}
\ket{j_1,j_2,j_3}=(-1)^{j_1-j_2+j_3}\sum_{m_1,m_2,m_3}  \left( \begin{array}{ccc}
\!\!j_1&j_2& j_3\!\!\\
\!\!m_1&m_2& m_3\!\! 
\end{array} \right)\ket{j_1,m_1}\ket{j_2,m_2}\ket{j_3,m_3}
\end{equation}
\item Expression for $6j$ coefficients
 \begin{equation}
\left\{ \begin{array}{ccc}
\!\!j_1&j_2& j_3\!\!\\
\!\!j_4&j_5& j_6\!\! 
\end{array} \right\}=\sum_{\text{all }p}{\cal P}\left( \begin{array}{ccc}
\!\!j_1&j_2& j_3\!\!\\
\!\!p_1&p_2& p_3\!\! 
\end{array} \right)\left( \begin{array}{ccc}
\!\!j_1&j_5& j_6\!\!\\
\!\!p_1&p_5& -p_6\!\! 
\end{array} \right)\left( \begin{array}{ccc}
\!\!j_4&j_2& j_6\!\!\\
\!\!-p_4&p_2& p_6\!\! 
\end{array} \right)\left( \begin{array}{ccc}
\!\!j_4&j_5& j_3\!\!\\
\!\!p_4&-p_5& p_3\!\! 
\end{array} \right)\nonumber~~~~~~~
\end{equation}
\end{enumerate}
Thus the  amplitudes for the gauge invariant magnetic field ground state  in the spin network basis on a tetrahedron are 
\begin{equation}
\Phi\left (\vec J\;\right)={\Pi(j_1,j_2,j_3,j_4,j_5,j_6)}\left\{ \begin{array}{ccc}
\!\!j_1&j_2& j_3\!\!\\
\!\!j_4&j_5& j_6\!\! 
\end{array} \right\}
\end{equation}
Thus using the pure gauge conditions and then integrating over all gauge degrees of freedom we can construct the magnetic ground states with ${\cal B}_p=0$). For lattice with non trivial geometry as torus we have more general solutions of ${\cal B}_p=0$ which generate
  degenerate topological ground states. On a 2 dimensional  lattice on torus we can draw two distinct non-contractible loops which can not be deformed into each other. We modify pure gauge conditions (\ref{pgc}) by $Z_2$ factors without changing the  magnetic fields as follows \cite{manatul}:
\begin{align}
\begin{aligned}
{Z}_{ab}=&\;\xi_a \,\eta_{\mathsf{q}}\,\xi^\dagger_b, ~~{Z}_{dc}=\;\xi_d \,\eta_{\mathsf{q}}\,\xi^\dagger_c,~~{Z}_{bc}=\;\xi_b \,\eta_{\mathsf{p}}\,\xi^\dagger_c, ~~{Z}_{ad}=\;\xi_a \,\eta_{\mathsf{p}}\,\xi^\dagger_d,\\
{Z}_{ba}=&\;\xi_b \,\xi^\dagger_a,~~~\,~~
{Z}_{cd}=\;\xi_c \,\xi^\dagger_d,~~~~~\,{Z}_{cb}=\;\xi_c \,\xi^\dagger_b,
~~~~\;{Z}_{da}=\;\xi_d\,\xi^\dagger_a. ~~~~~~~
\end{aligned}
\end{align}
Where $\eta_{\mathsf{p}}=e^{i2\pi \mathsf{p}}= (-1)^{2\mathsf{p}}$, $\eta_{\mathsf{q}}=e^{i2\pi \mathsf{q}}= (-1)^{2\mathsf{q}}$ with $ \mathsf{p}, \mathsf{q}=0$. 
With this solution of ordered states we can integrates all the gauge degrees of freedom to get amplitudes for topological ground states:
\begin{align*}
\Phi_{({\mathsf p},\,{\mathsf q})}\left(\vec J_a,\vec J_b,\vec J_c,\vec J_d\right)
=& (-1)^{2 {\mathsf p} \left(j_2^a+j_2^b\right)} ~(-1)^{2 {\mathsf q}\left(j_1^a+j_1^d\right)} ~~ \\
&~~~~~~~~~\times\Pi (\{ \vec{J}\})\begin{bmatrix}\; j_2^a=j_4^d\!\! && \!\! j_1^a=j_3^b\!\! && \!\! j_2^c=j_4^b\!\! && \!\! j_{1}^c=j_3^d\!\! &&\\
&\!\! j^a \!\! && \!\! j^b \!\! && \!\! j^c\!\! && \!\! j^d\!\! \!\! \!\! \\
\;j_2^d =j_4^a\!\! &&\!\! j_1^b =j_3^a\!\! && \!\! j_2^b =j_4^c\!\! && \!\! j_1^d=j_3^c\!\! &&\\
\end{bmatrix}
\end{align*}
\section{Matrix Elements}
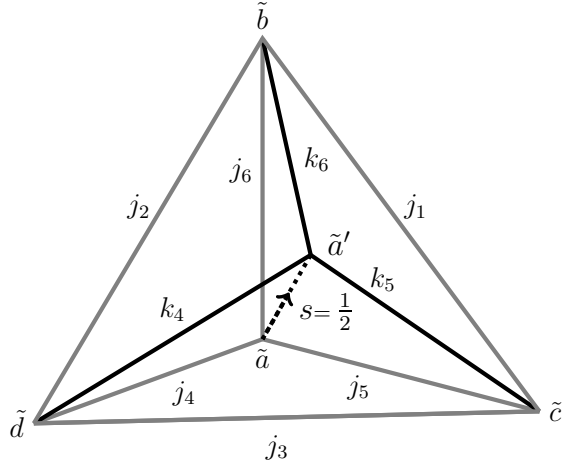
\begin{figure}
	\begin{center}
	\begin{tikzpicture}[scale=1.6]
\coordinate (a) at (-1.9,-.7);
\coordinate (b) at (2.3,-.6);
\coordinate (c) at (0,0);
\coordinate (d) at (0,2.5);
\coordinate (o) at (.4,.7);
\draw[gray,ultra thick ] (a)--(b);
\draw[gray,ultra thick ] (a)--(c);
\draw[gray, ultra thick] (c)--(b);
\draw[ultra thick,gray] (d)--(c);
\draw[ultra thick] (a)--(o)--(b);
\draw[ultra thick] (d)--(o);
\draw[ultra thick,gray] (a)--(d)--(b);
\draw[ultra thick,gray] (a)--(b);
\draw[dotted, ultra thick] (o)--node[pos=.3](x){}(c);
\draw[dotted, ultra thick,->] (c)--(x);
\node[scale=.9,left] at (a) {$\tilde{d}$};
\node[scale=.9,right] at (b) {$\tilde{c}$};
\node[scale=.9,below] at (c) {$\tilde{a}$};
\node[scale=.9,above] at (d) {$\tilde{b}$};
\node[scale=.9] at (-.65,-.45) {$j_4$};
\node[scale=.9] at (.8,-.4) {$j_5$};
\node[left,scale=.9] at (0,1.4) {$j_6$};
\node[left,scale=.9] at (0.3,-.9) {$j_3$};
\node[left,scale=.9] at (1.45,1.1) {$j_1$};
\node[left,scale=.9] at (-.85,1.1) {$j_2$};
\node[scale=.9] at (-.75,0.25) {$k_4$};
\node[scale=.9] at (1.,0.5) {$k_5$};
\node[scale=.9] at (.45,1.5) {$k_6$};
\node[] at (.53,.23) {$s{\scriptstyle =}\,\tfrac{1}{2}$};
\node[right] at (.45,.8) {$\tilde{a}'$};
	\end{tikzpicture}
\end{center}
\caption{The simple action of the operator Tr $\triangle_{bcd}$ on the dual lattice. It moves the single dual site: $\tilde a \rightarrow {\tilde a}'$. The matrix elements  $\mathcal{M}^{(bcd)}$ are 9j coefficients in (\ref{iwt}).}\label{cal9j}
\end{figure}

In this section we will calculate matrix elements of various  Wilson loop operators in spin network basis. The calculations are done for a triangular loop ${\cal W}_{\cal C} =\triangle_{bcd}$ 
in the spin $s= \frac{1}{2}$ representation. We then generalise the results to other cases discussed in this work. The various triangular constraints and the matrix elements are displayed in the dual tetrahedron  diagram in the angular momentum space in Figure \ref{cal9j}.  All angular momenta meeting at a vertex in real space in Figure \ref{tetra6j} now lie on a triangle in the dual angular momentum space in Figure \ref{cal9j}. The reverse is also true. Therefore, the action of the Tr $\triangle_{bcd}$ is to move the single dual site
${\tilde a} \rightarrow {\tilde a}'$ this is shown in Figure \ref{cal9j} by a dotted line with an arrow. As Wilson loops create or destroy half unit of flux on the links so they act like a ladder or creation and annihilation operators for the SU(2) fluxes in the spin network states. The matrix element $ \bra{\{\vec J\}} \;\triangle_{bcd}\; \ket{\{\vec K\}}$ will be non-zero only if $\{k=j\pm\frac{1}{2}\}$ along the Wilson loop, i.e $l\in {\cal C}$. The flux values on all other links,  $l \notin {\cal C}$, remain unaltered i.e.,  $k=j$.  Moreover there are new triangular constraints between  $\vec J $ and $\vec K$ and $s=\frac{1}{2}$ which are clearly shown in Figure \ref{cal9j} which represents a 9j Wigner coefficients. 
\bea 
{\cal M}^{(bcd)}_{[\{\vec J\}~\{\vec K\}]}  =\frac{1}{2}  \bra{\{\vec J\}} \;\triangle_{bcd}\; \ket{\{\vec K\}}\hspace{5cm} \nonumber \\ 
=  
\frac{1}{2}\delta_{j_1,\, k_1}\delta_{j_2,\, k_2}
\delta_{j_3,\, k_3}~\Pi( j_4,\,k_4,j_5,\,k_5,\,j_6.\,k_6)
\left\{ \begin{array}{cccc}
{j}_{4} & {k}_{4} & 1/2 \\
{k}_{5} & j_{5} & j_3\\
\end{array} \right \}
\left\{ \begin{array}{cccc}
{j}_{5} & {k}_{5} & 1/2 \\
{k}_{6} & j_{6} & j_1\\
\end{array} \right \}
\left\{ \begin{array}{cccc}
{j}_{6} & {k}_{6} & 1/2 \\
{k}_{4} & j_{4} & j_2\\
\end{array} \right \}\\ \nonumber \\
=  \frac{M^{bcd}}{4} \sum_{x}(-1)^{R+2x}(2x+1)
\left\{ \begin{array}{cccc}
{j}_{4} & {k}_{4} & x \\
{k}_{5} & j_{5} & j_3\\
\end{array} \right \}
\left\{ \begin{array}{cccc}
{j}_{5} & {k}_{5} & x\\
{k}_{6} & j_{6} & j_1\\
\end{array} \right \}
\left\{ \begin{array}{cccc}
{j}_{6} & {k}_{6} & x \\
{k}_{4} & j_{4} & j_2\\
\end{array} \right \} 
\prod_{l=4,5,6}\left\{ j_l\; k_l \;\frac{1}{2}\right\}
 ~~~~\nonumber \\
 = \frac{M^{bcd}}{\Pi^4(\frac{1}{2})}~\left[\begin{array}{cccccc}
\!\!{k}_{4}\!\! & & \!\!{k}_{5}\!\! &  &\!\! k_6\!\! &\\
& \!\!j_3\!\! & & \!\!j_1\!\! & &\!\! j_2\!\!\!\\  
\!\!{j}_{4} \!\!&   & \!\!{j}_{5} \!\!&  & \!\!j_6 \!\!& \!\!\\
\end{array} \right]\,\prod_{l=4,5,6}\left\{ j_l,\; k_l, \;\frac{1}{2}\right\} \hspace{5cm}
\label{iwt}
\eea
Here 
\begin{align*}
M^{bcd}= \delta_{j_1,\, k_1}\delta_{j_2,\, k_2}\delta_{j_3,\, k_3} \Pi( j_4,\,k_4,j_5,\,k_5,\,j_6.\,k_6). 
\end{align*}
It is easy to verify the validity of (\ref{iwt}). 
The 9j matrix structure is uniquely fixed by the triangular constraints of the spin networks.
The numerical factor $M^{bcd}$  can be verified  by replacing $ \frac{\triangle_{bcd}}{2}$  by an identity operators and 
getting  ${\cal M}^{bcd}=\langle \;\{\vec J\} \;| \;\{\vec K\} \;\rangle = \delta_{\vec J,\,\vec K}$. The matrix elements (\ref{iwt}) are geometrical in nature and therefore can be easily generalize to other cases discussed in this work. The associated numerical factors can also be checked by replacing the Wilson loop operator by an identity operator as done in the present case.
\section{Proof of the 6j-9j identity} \label{proof} 
In this section, using the known results in the literature,  we  prove the  identities  derived in section \ref{tetra}. 
We use the following two identities from Varshalovich \cite{varsha}: 
\begin{enumerate} 
\item We use  the standard expansion of 9j in terms of 6j 
given in \cite{varsha} on page 361, equation (2) for n=3. 
\begin{equation}
\left[\begin{array}{cccccc}
\!j_1\!\! & & \!\!{j}_2 \!\!&  & \!\!j_3\!\!&\!\!\\
&\!\!l_1\!\!& &\!\!l_2\!\!& &\!\! l_3\!\!\!\\  
\!k_1\!\! & & \!\!{k}_2 \!\!&  & \!\!k_3\!\!&\!\!\\
\end{array} \right]=\sum_{x} \Pi^2(x) (-1)^{R+3x} \left\{ \begin{array}{cccc}
\!\!{j}_{1} & \!{k}_{1} \!&\! x \!\!\!\\
\!\!{k}_{2} &\! j_{2} \!&\! l_1\!\!\!\\
\end{array} \right \}\left\{ \begin{array}{cccc}
\!\!{j}_{2} & \!{k}_{2} \!&\! x \!\!\!\\
\!\!{k}_{3} &\! j_{3} \!&\! l_2\!\!\!\\
\end{array} \right \}\left\{ \begin{array}{cccc}
\!\!{j}_{3} & \!{k}_{3} \!&\! x \!\!\!\\
\!\!{k}_{1} &\! j_{1} \!&\! l_3\!\!\!\\
\end{array} \right \}.  ~~\text{(V-1)}
\label{varsha1}  \nonumber 
\end{equation}
Here  $R=j_1+j_2+j_3+ l_1+l_2+l_3+k_1+k_2+k_3$. 
\item The second identity involves triple summations and 
it is given on page 472, equation (38):  
\begin{align}
\begin{aligned}\label{varsha2}
\sum_{k_1,\,k_2,\,k_3} \Pi^2(k_1,k_2,k_3) (-1)^{R+x} \left\{ \begin{array}{cccc}
\!\!{j}_{1} & \!{k}_{1} \!&\! x \!\!\!\\
\!\!{k}_{2} &\! j_{2} \!&\! l_3\!\!\!\\
\end{array} \right \}\left\{ \begin{array}{cccc}
\!\!{j}_{2} & \!{k}_{2} \!&\! x \!\!\!\\
\!\!{k}_{3} &\! j_{3} \!&\! l_1\!\!\!\\
\end{array} \right \}&\left\{ \begin{array}{cccc}
\!\!{j}_{3} & \!{k}_{3} \!&\! x \!\!\!\\
\!\!{k}_{1} &\! j_{1} \!&\! l_2\!\!\!\\
\end{array} \right \}\left\{ \begin{array}{cccc}
\!\!{l}_{1} & \!{l}_{2} \!&\! l_3 \!\!\!\\
\!\!{k}_{1} &\! k_{2} \!&\! k_3\!\!\!\\
\end{array} \right \}\\
&= \Pi^2(x) \left\{ \begin{array}{cccc}
\!\!{l}_{1} & \!{l}_{2} \!&\! l_3 \!\!\!\\
\!\!{j}_{1} &\! j_{2} \!&\! j_3\!\!\!\\
\end{array} \right \}.   ~~~~~~~\text{(V-2)} \nonumber 
\end{aligned}
\end{align} 
 \end{enumerate} 
Now we consider left hand side of the identity (\ref{itype11})
\begin{align}
\hspace{-2.5cm} 
\begin{aligned}
\text{LHS}=&\;\sum_{\{k\}'} 
\Pi^2(k_4,k_5,k_6) 
 \left[\begin{array}{cccccc}
\!\!{k}_{4}\!\! & & \!\!{k}_{5}\!\! &  &\!\! k_6\!\! &\\
& \!\!j_3\!\! & & \!\!j_1\!\! & &\!\! j_2\!\!\!\\  
\!\!{j}_{4} \!\!&   & \!\!{j}_{5} \!\!&  & \!\!j_6 \!\!& \!\!\\
\end{array} \right]\left\{ \begin{array}{cccc}
{j}_{1} & {j}_{2} & j_3 \\
{k}_{4} & k_{5} & k_6\\
\end{array} \right\}\\
=&\;\sum_{k_4,k_5,k_6} 
\Pi^2(k_4,k_5,k_6) 
\left[\begin{array}{cccccc}
\!\!{k}_{4}\!\! & & \!\!{k}_{5}\!\! &  &\!\! k_6\!\! &\\
& \!\!j_3\!\! & & \!\!j_1\!\! & &\!\! j_2\!\!\!\\  
\!\!{j}_{4} \!\!&   & \!\!{j}_{5} \!\!&  & \!\!j_6 \!\!& \!\!\\
\end{array} \right]\left\{ \begin{array}{cccc}
{j}_{1} & {j}_{2} & j_3 \\
{k}_{4} & k_{5} & k_6\\
\end{array} \right\}\prod_{l=4,5,6}\{ j_l, k_l , s\} \nonumber
\end{aligned}
\end{align} 
We expand $9j$ coefficients using (V-1):
\begin{align}
\hspace{-1cm} 
\begin{aligned}
\text{LHS}=&\sum_{k_4,\;k_5,\;k_6} 
\Pi^2(k_4,k_5,k_6) 
\sum_{x} (-1)^{R+x} \Pi^2(x)
 \left\{ \begin{array}{cccc}
{k}_{4} & {j}_{4} & x \\
{j}_{5} & k_{5} & j_3\\
\end{array} \right\}
 \left\{ \begin{array}{cccc}
{k}_{5} & {j}_{5} & x \\
{j}_{6} & k_{6} & j_1\\
\end{array} \right\}\\
&\qquad\qquad\qquad\qquad\qquad\qquad\qquad \times\left\{ \begin{array}{cccc}
{k}_{6} & {j}_{6} & x \\
{j}_{4} & k_{4} & j_3\\
\end{array} \right\}\left\{ \begin{array}{cccc}
{j}_{1} & {j}_{2} & j_3 \\
{k}_{4} & k_{5} & k_6\\
\end{array} \right\}\prod_{l=4,5,6}\{ j_l, k_l , s\}\nonumber
\end{aligned}
\end{align} 
In above $R=j_1+j_2+j_3+ j_4+j_5+j_6+k_4+k_5+k_6$. Due to triangular constraints $\{j_l\; k_l, s\}$  only non-zero term in the series is $x=s$; i. e.
\begin{align}
\begin{aligned}
\hspace{-4cm}&\text{LHS}\\
&=\!\Pi^2(s)\!\!\!\sum_{k_4,\,k_5,\,k_6}\!\!(-1)^{R+s}\;
\Pi^2(k_4,k_5,k_6) 
 \left\{ \begin{array}{cccc}
\!\!{k}_{4} & {j}_{4} & s\!\! \\
\!\!{j}_{5} & k_{5} & j_3\!\!\\
\end{array} \right\}
 \left\{ \begin{array}{cccc}
\!\!{k}_{5} & {j}_{5} & s \!\!\\
\!\!{j}_{6} & k_{6} & j_1\!\!\\
\end{array} \right\}\left\{ \begin{array}{cccc}
\!\!{k}_{6} & {j}_{6} & s \!\!\\
\!\!{j}_{4} & k_{4} & j_3\!\!\\
\end{array} \right\}\left\{ \begin{array}{cccc}
\!\!{j}_{1} & {j}_{2} & j_3 \!\!\\
\!\!{k}_{4} & k_{5} & k_6\!\!\\
\end{array} \right\}\\
&=\Pi^4(s)\left\{ \begin{array}{cccc}
\!\!{j}_{1} & {j}_{2} & j_3 \!\!\\
\!\!{j}_{4} & j_{5} & j_6\!\!\\
\end{array} \right\}~~~~~~~~~~~~~~~~~~~~~~~~~~~~~ \text{( using identity (V-2) )}\\
&=\text{RHS} \nonumber
\end{aligned}
\end{align} 

\end{document}